\documentclass[11pt,a4paper]{article}
\usepackage[margin=25mm]{geometry}
\usepackage[T1]{fontenc}
\usepackage{lmodern}
\usepackage[stretch=30,shrink=30]{microtype}
\usepackage{graphicx,placeins}
\usepackage[numbers,sort&compress]{natbib}
\usepackage{amsmath}
\usepackage{amssymb}
\usepackage{bm}
\usepackage{booktabs}
\usepackage{array}
\usepackage{siunitx}
\usepackage{subcaption}
\usepackage[hidelinks]{hyperref}
\newcolumntype{L}[1]{>{\raggedright\arraybackslash}p{#1}}
\newcolumntype{C}[1]{>{\centering\arraybackslash}p{#1}}
\newcolumntype{R}[1]{>{\raggedleft\arraybackslash}p{#1}}
\graphicspath{{figures/}}
\AddToHook{env/figure/begin}{\linespread{1.0}\selectfont}
\AddToHook{env/table/begin}{\linespread{1.0}\selectfont\setlength{\abovecaptionskip}{0pt}\setlength{\belowcaptionskip}{6pt}}
\AddToHook{env/tabular/begin}{\footnotesize\linespread{1.0}\selectfont}

\begin{document}

\title{Localization Reversal Before and After Uncoupled Mode Crossings in Phononic Crystal Defect Dimers}

\author{Xinlin Xu\thanks{Corresponding author. Email: \href{mailto:xu.xinlin.k4@f.mail.nagoya-u.ac.jp}{\texttt{xu.xinlin.k4@f.mail.nagoya-u.ac.jp}}}\quad Junji Kato\\[0.5em]
\small Department of Civil and Environmental Engineering,\\
\small Graduate School of Engineering, Nagoya University,\\
\small Furo-cho, Chikusa-ku, Nagoya, Aichi 464-8601, Japan}
\date{}
\maketitle

\begin{abstract}
Softening one defect in a phononic-crystal dimer can draw a localized elastic mode toward it and then shift the same mode to the unchanged defect. We study when this switch occurs relative to the frequency crossing of two uncoupled defect modes. Each uncoupled mode spans both sites, with a site bias given by the fraction of its kinetic energy in the unchanged cell minus that in the softened cell. At the crossing, the two modes enter the coupled two-state eigenvectors with equal coefficient magnitudes. If the summed site biases exceed the interference contribution in magnitude, both eigenvectors favor the site indicated by that sum. For branches that switch sites once, this criterion places reversal before the crossing in a poly(methyl methacrylate)--aluminum dimer and after it in an aluminum--tungsten dimer, in agreement with full-order calculations. At reversal in aluminum--tungsten, the displacement projected onto the crossing pair still favors the softened defect despite retaining 92.17\% of the modal mass. Interference with the remaining displacement of the same mirror symmetry supplies most of the energy difference needed to balance the two cells. A multimode model using only the nominal structure and prescribed stiffness change predicts both reversal thresholds within 1\% of the interpolated full-order values.
\end{abstract}

\noindent\textbf{Keywords:} Phononic crystals; elastic waves; defect modes; mode localization; mode hybridization; avoided crossings

\section{Introduction}
\label{sec:introduction}

\begin{figure}[!b]
  \centering
  \begin{minipage}[t]{0.48\linewidth}
    \centering\includegraphics[trim=0 8pt 0 0,clip,width=\linewidth,height=142pt,keepaspectratio]{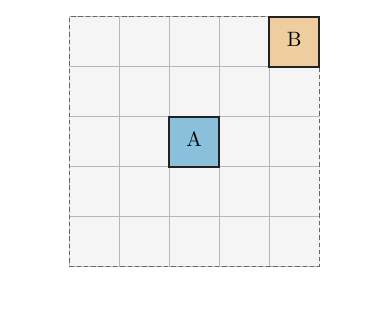}
    \par\smallskip{\small (a) Diagonal dimer layout.\par}
  \end{minipage}\hfill
  \begin{minipage}[t]{0.48\linewidth}
    \centering\includegraphics[trim=0 8pt 0 0,clip,width=\linewidth,height=142pt,keepaspectratio]{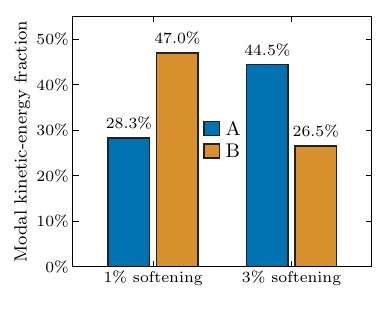}
    \par\smallskip{\small (b) Defect-cell kinetic energies.\par}
  \end{minipage}
  \caption{Localization reversal in a periodic aluminum--tungsten supercell. The bars give the fractions of total cycle-averaged modal kinetic energy in the unchanged defect $A$ and softened defect $B$ on the same full-order branch at 1\% and 3\% softening.}
  \label{fig:opening}
\end{figure}

A point defect can trap an elastic wave at a frequency inside the bandgap of a phononic crystal~\citep{Mead1973,Kushwaha1993,Khelif2003Defect,Hussein2014}. With two nearby defects, changing the stiffness of one site can shift the vibration between them. Under progressive softening the same mode can first concentrate at the softened site and then at its unchanged partner (Fig.~\ref{fig:opening}). Two spectral branches can exchange character near an avoided crossing, but each participating mode may carry energy at both sites. The question is then whether the vibration has already switched sites when the underlying uncoupled frequencies meet.

Mode localization and eigenvalue veering in nearly periodic structures are closely related~\citep{Hodges1982,Pierre1988Localization}. Disorder in layer thickness or elastic constants, for example, can localize Rayleigh waves in piezoelectric phononic crystals~\citep{Wang2008RayleighLocalization}. Eigenvector sensitivity explains why close frequencies can accompany large changes in mode shape~\citep{PerkinsMote1986,Triantafyllou1991,Liu2002EigenSensitivity}. Distinguishing avoided crossings from true degeneracies is therefore important when tracking phononic bands~\citep{LuSrivastava2018BandSorting}, as illustrated by the distinct crossings in plates with tapered resonators~\citep{Das2023AvoidedCrossingsJAP}. Parameter gradients and coupling can alter localization~\citep{BenLassoued2023Localization}, and nearby modes interact in inertially coupled oscillators~\citep{JacquesAdhikari2025} and imperfect shells~\citep{Kim2026ShellLocalization}. For a defect pair, the change in modal composition must be related to a spatial observable: the kinetic energy in each defect cell.

Defect-localization models already predict more than eigenfrequencies. Exact analyses of bi-periodic mass--spring systems describe single-disorder localization in one and two dimensions~\citep{Cai2003SingleDisorder,Cai2005SingleDisorder}, while continuum-supercell studies identify effects of defect location and cell size~\citep{Jo2020Supercell}. Ramakrishnan and Matlack~\citep{RamakrishnanMatlack2025} quantitatively predict defect frequencies, mode shapes, and localized amplitude envelopes in monoatomic phononic crystals; their multiple-defect extension applies when the defects lie outside one another's proximal influence zones. Multiple localized modes can also coexist at a single defect~\citep{Marchal2012}.

Coupling between neighboring defects can split a defect band into two resonances~\citep{LiLiu2005,RomeroGarcia2010DoubleDefects,Jo2020DoubleDefectJAP} and supports coupled-resonator waveguides and acoustic circuits~\citep{Escalante2013,Ma2022CoupledResonators}. In an aluminum plate, Jo et al.~\citep{Jo2021} resolved split bands and in-phase and out-of-phase modes derived from one monopole-like defect resonance. Tuning differently patterned defects can also change the site of energy concentration~\citep{Jo2022Localization,Jo2023DoubleDefects}. Related two-site modes occur in soft-clamped mechanical dimers~\citep{Catalini2020}. Two localized copies of one resonance provide a natural basis for an isolated modal family. When several defect modes are close in frequency, symmetry and overlap govern which modes can mix, as in coupled degenerate photonic defects~\citep{Mahmoodian2009DegenerateDefects}. Multimode coupled-mode theory accounts for the resulting energy normalization and interference~\citep{Fan2003,Suh2004}. In elastic lattices, continuum models resolve modal interactions at higher frequencies~\citep{Telgen2024RainbowTrapping}, and Bloch eigenmodes provide a basis for effective elastodynamic descriptions~\citep{Sridhar2018EffectiveElastodynamics}.

A synthetic negative capacitance can tune the resonance frequency of a single piezoelectric defect~\citep{Jo2024SNCFilterJAP}, while electrical tuning of a defect pair can vary the coupling between the two defects~\citep{Jo2024TunableCoupling}. Segmented piezoelectric elements and feedback can control multiple defect bands~\citep{Jo2025Steering,Jo2026Feedback}, and inverse design can target frequency and mode shape together~\citep{Jo2026FrequencyModeMatching}. Magnetic loading in soft phononic-crystal beams~\citep{ZhangZhang2025MagneticDefects} and graded electrical resonances in piezoelectric metamaterial beams~\citep{Alshaqaq2027Localization} likewise tune defect frequencies or spatial localization profiles. When a dimer supports several nearby mode families, however, altering the stiffness of a single defect can shift the balance of vibration between the two sites in ways that single-family descriptions do not capture.

Zhang et al.~\citep{Zhang2017ForcedLocalization} showed that, in weakly coupled micromechanical resonators under double-resonator forcing, two resonances can contribute equally to the response away from the veering point. For a free vibration mode, the kinetic energy in each defect cell follows from the mass-weighted squared displacement integrated over that cell.

Here we follow one localized mode in each of two diagonal elastic dimers while softening a single defect at fixed geometry and density. The mode switches sites before the uncoupled frequency crossing in one dimer and after it in the other. We derive the preferred site at the crossing from the signed kinetic-energy differences of the two uncoupled modes, then test how much of the coupled displacement is needed to locate the switch. In aluminum--tungsten the crossing pair still favors the softened site at the reversal threshold. Resolving its interference with the remaining displacement explains how the two cells nevertheless reach equal kinetic energy. Independent finite-element solutions test the resulting threshold predictions, while axial and off-axis layouts isolate the role of symmetry in coupling the defect modes.

\section{Elastic Dimers and Measures of Localization}
\label{sec:numerical-setup}

To examine how stiffness detuning redistributes vibration between neighboring defects, we consider a periodic supercell containing two copies of the same defect. In each fixed layout, only the Young's modulus of defect $B$ is varied, while geometry and mass remain unchanged. The eigenmodes are followed along detuning branches, and localization is evaluated from the physical modal mass in the two defect cells. Comparing coupled and independent-channel descriptions under this common localization measure identifies the interactions needed to capture the spatial response.

\subsection{PMMA--Aluminum Single-Defect System}

\begin{figure}[!b]
  \centering
  \begin{minipage}[t]{0.48\linewidth}
    \centering
    \includegraphics[width=\linewidth,height=150pt,keepaspectratio]{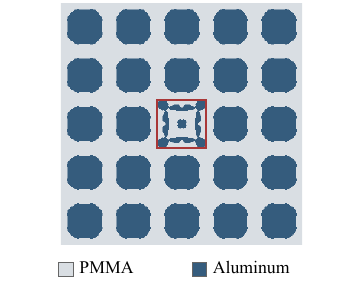}
    \par\smallskip{\small (a) Single-defect reference.\par}
  \end{minipage}
  \hfill
  \begin{minipage}[t]{0.48\linewidth}
    \centering
    \includegraphics[width=\linewidth,height=150pt,keepaspectratio]{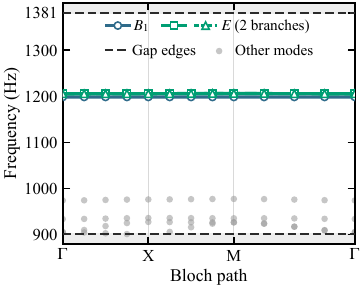}
    \par\smallskip{\small (b) Dispersion spectrum.\par}
  \end{minipage}
  \par\medskip
  \begin{minipage}[t]{0.30\linewidth}
    \centering
    \includegraphics[width=\linewidth,height=90pt,keepaspectratio]{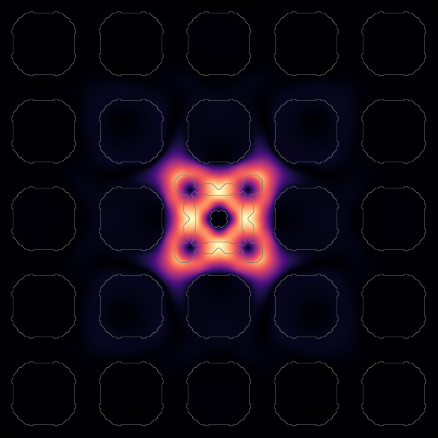}
    \par\smallskip{\small (c) $B_1$, \SI{1198.751}{Hz}.\par}
  \end{minipage}
  \hfill
  \begin{minipage}[t]{0.30\linewidth}
    \centering
    \includegraphics[width=\linewidth,height=90pt,keepaspectratio]{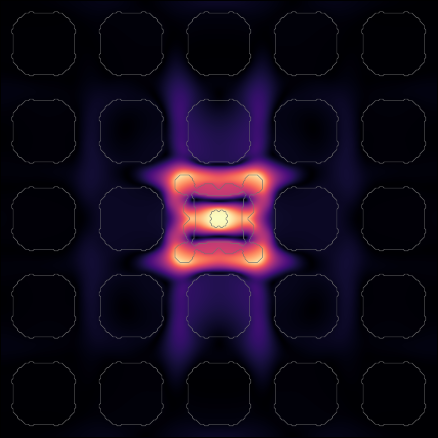}
    \par\smallskip{\small (d) $E$, $\sigma_x$ even, \SI{1205.721}{Hz}.\par}
  \end{minipage}
  \hfill
  \begin{minipage}[t]{0.30\linewidth}
    \centering
    \includegraphics[width=\linewidth,height=90pt,keepaspectratio]{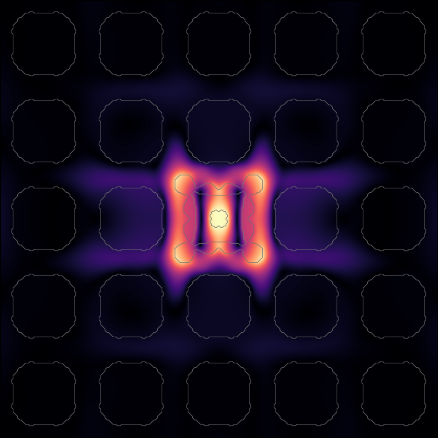}
    \par\smallskip{\small (e) $E$, $\sigma_x$ odd, \SI{1205.721}{Hz}.\par}
  \end{minipage}
  \par\medskip
  \includegraphics[width=190bp]{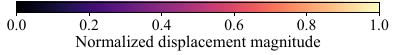}
\caption{PMMA--aluminum system: single-defect structure, dispersion, and localized modes. The two $E$ branches overlap; gray markers denote other in-gap modes. Here $\Gamma$, $\mathrm X$, and $\mathrm M$ are the center, an edge midpoint, and a corner of the supercell Brillouin zone. Displacement fields are normalized independently for display; thin contours mark material interfaces.}
  \label{fig:fixed-study-models}
\end{figure}

The first system is a two-phase poly(methyl methacrylate) (PMMA)--aluminum phononic crystal with a $C_{4v}$-symmetric defect in a $5\times5$ periodic supercell (Fig.~\ref{fig:fixed-study-models}). The single defect supports a nondegenerate $B_1$ mode (singlet) and a degenerate $E$ mode pair (doublet) with nearby frequencies inside the host bandgap. The labels identify one- and two-dimensional irreducible representations of the single-defect point group $C_{4v}$, respectively. The doublet partners are even and odd under reflection $\sigma_x$ about the $x$ axis. The lattice constant is $a=\SI{1}{m}$, and the material properties are listed in Table~\ref{tab:numerical-setup}.

\begin{table}[!b]
  \centering
  \caption{Material properties of the PMMA--aluminum system.}
  \label{tab:numerical-setup}
  \setlength{\tabcolsep}{7pt}
  \begin{tabular}{clS[group-digits=none,table-format=2.1]S[group-digits=none,table-format=4.0]S[group-digits=none,table-format=1.2]}
    \toprule
    Phase & Material & {$E$ (\si{GPa})} & {$\rho$ (\si{kg.m^{-3}})} & {$\nu$} \\
    \midrule
    1 & PMMA & 3.0 & 1180 & 0.35 \\
    2 & Aluminum & 70.0 & 2700 & 0.35 \\
    \bottomrule
  \end{tabular}
\end{table}

\subsection{Aluminum--Tungsten Single-Defect System}
\label{sec:al-w-dimer}

In the diagonal PMMA--aluminum dimer, the $B_1$ singlet couples to the component of the $E$ doublet that is odd under reflection about the inter-defect diagonal. To test reversal with the other mirror parity, we consider an aluminum--tungsten crystal whose defect supports a $B_2$ singlet. This singlet and the even component of $E$ share a mirror sector in the diagonal dimer.

\begin{figure}[!t]
  \centering
  \begin{minipage}[t]{0.48\linewidth}
    \centering
    \parbox[c][150pt][c]{\linewidth}{\centering\includegraphics[width=\linewidth,height=150pt,keepaspectratio]{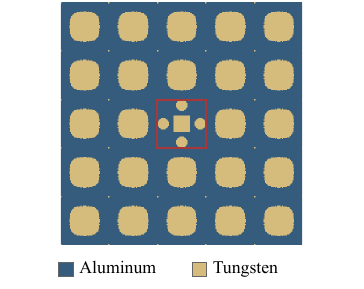}}
    \par\smallskip{\small (a) Single-defect reference.\par}
  \end{minipage}\hfill
  \begin{minipage}[t]{0.48\linewidth}
    \centering
    \parbox[c][150pt][c]{\linewidth}{\centering\includegraphics[width=\linewidth,height=150pt,keepaspectratio]{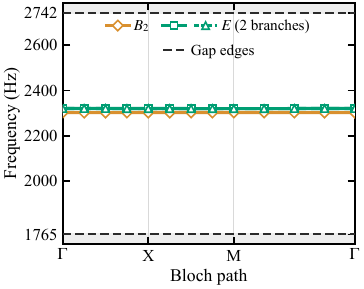}}
    \par\smallskip{\small (b) Dispersion spectrum.\par}
  \end{minipage}
  \caption{Aluminum--tungsten system: single-defect structure and dispersion. In (a), the red box marks the central defect cell in the $5\times5$ periodic supercell. In (b), the full-order single-defect spectrum shows the $B_2$ branch and the two branches ($E_1,E_2$) continued from the $E$ doublet along $\Gamma$--$\mathrm X$--$\mathrm M$--$\Gamma$. The two $E$ branches overlap at this scale.}
  \label{fig:al-w-geometry}
\end{figure}

Figure~\ref{fig:al-w-geometry}(a) shows the single-defect supercell, with material properties listed in Table~\ref{tab:al-w-materials}. The defect consists of a tungsten square of side $a/3$ at the center of an aluminum cell and four tungsten disks of radius $7a/60$, centered at $(\pm23a/60,0)$ and $(0,\pm23a/60)$ relative to the cell center, preserving $C_{4v}$ symmetry.

\begin{table}[!t]
  \centering
  \caption{Material properties of the aluminum--tungsten system.}
  \label{tab:al-w-materials}
  \setlength{\tabcolsep}{7pt}
  \begin{tabular}{clS[group-digits=none,table-format=3.1]S[group-digits=none,table-format=5.0]S[group-digits=none,table-format=1.2]}
    \toprule
    Phase & Material & {$E$ (\si{GPa})} & {$\rho$ (\si{kg.m^{-3}})} & {$\nu$} \\
    \midrule
    1 & Aluminum & 70.0 & 2700 & 0.35 \\
    2 & Tungsten & 400.0 & 19300 & 0.35 \\
    \bottomrule
  \end{tabular}
\end{table}

At $\Gamma$ [Fig.~\ref{fig:al-w-geometry}(b)], the single-defect $B_2$ and $E$ frequencies are \SI{2301.776}{Hz} and \SI{2319.792}{Hz}. All three branches lie inside the host bandgap of \SIrange{1764.555}{2741.534}{Hz} along the sampled path. At the 13 path points, the minimum distance to either gap edge is \SI{421.66}{Hz}, and the defect cell carries at least 66.48\% of the modal mass.

The square size, disk radius, and disk position specified above were selected with the six-coordinate nominal model so that an uncoupled $B_2$--$E$ crossing and a predicted substantial reversal occur within the softening interval $-0.03<\alpha<0$. We then tested that nominal prediction against independent full-order calculations.

\subsection{Dimer Layouts and Stiffness Detuning}

\begin{figure}[!b]
  \centering
  \begin{minipage}[t]{0.32\linewidth}
    \centering
    \includegraphics[width=\linewidth,height=106pt,keepaspectratio]{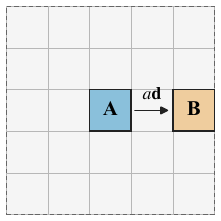}
    \par\smallskip{\small (a) Axial, $\bm{d}=(2,0)$\newline (PMMA--aluminum).\par}
  \end{minipage}
  \hfill
  \begin{minipage}[t]{0.32\linewidth}
    \centering
    \includegraphics[width=\linewidth,height=106pt,keepaspectratio]{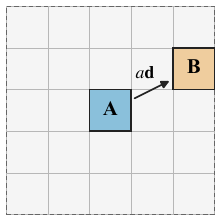}
    \par\smallskip{\small (b) Off-axis, $\bm{d}=(2,1)$\newline (PMMA--aluminum).\par}
  \end{minipage}
  \hfill
  \begin{minipage}[t]{0.32\linewidth}
    \centering
    \includegraphics[width=\linewidth,height=106pt,keepaspectratio]{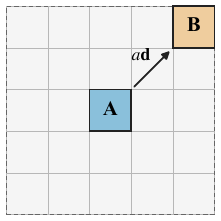}
    \par\smallskip{\small (c) Diagonal, $\bm{d}=(2,2)$\newline (both systems).\par}
  \end{minipage}
  \caption{Dimer layouts in the $5\times5$ periodic supercell. Defect $A$ (blue) remains unchanged at cell $(2,2)$, and defect $B$ (amber) is detuned at offset $a\bm d=\bm r_B-\bm r_A$ with separation $r=a\|\bm d\|_2$. The PMMA--aluminum dimer is studied in all three layouts, and the aluminum--tungsten dimer in the diagonal layout (c).}
  \label{fig:numerical-geometry-schematic}
\end{figure}

Both material systems use a $60\times60$ mesh per cell with plane-strain finite elements and consistent mass; the integration and solver settings are collected in Sec.~\ref{app:rom-settings} of the supplementary material. A dimer is formed by placing two identical copies of the single defect in the $5\times5$ periodic supercell, so the reported interactions include the periodic host response at $\Gamma$. Defect $A$ is fixed at cell $(2,2)$ and defect $B$ is placed at offset $\bm d=(2,0)$, $(2,1)$, or $(2,2)$, defining axial, off-axis, and diagonal layouts with separations $2a$, $\sqrt5a$, and $2\sqrt2a$ (Fig.~\ref{fig:numerical-geometry-schematic}). The PMMA--aluminum system is examined across all three layouts to assess how dimer symmetry and coupling govern localization, whereas the aluminum--tungsten system is studied in the diagonal layout [Fig.~\ref{fig:numerical-geometry-schematic}(c)] to compare the opposite mirror parity under the same dimer geometry. Within each layout, all models are compared for the same physical configuration and detuning.

The dimer comparisons use Bloch-periodic conditions at $\Gamma$ ($\bm k=\bm0$). The dimensionless detuning $\alpha$ changes the Young's modulus of both material phases throughout defect cell $B$ according to
\begin{equation}
  E_B(\alpha)=(1+\alpha)E_B(0).
  \label{eq:detuning}
\end{equation}

Here $\alpha=0$ is the nominal state; positive and negative values stiffen and soften $B$, respectively. Geometry, densities, Poisson's ratios, and properties outside $B$ remain fixed. Thus $\bm K(\alpha)=\bm K_0+\alpha\bm K_B$ with constant mass, where $\bm K_0$ is the nominal stiffness matrix and $\bm K_B$ is the stiffness contribution assembled from defect $B$. Plots that report detuning in percent use $100\alpha$ for signed detuning and $-100\alpha$ for softening, while other plots use dimensionless $\alpha$ or softening magnitude $-\alpha$. The PMMA--aluminum full-sweep accuracy comparisons use 61 equally spaced values in $[-0.03,0.03]$; the reversal controls in Sec.~\ref{sec:fixed-branch-results} follow the softening branch.

\subsection{Localization and Branch Identity}
\label{sec:mechanism-ablation}

The full-order model (FOM) solves the Bloch-reduced generalized eigenproblem~\citep{Mead1973,Hussein2014}
\begin{equation}
  \bm{K}(\bm{k})\bm{q}_r
  =\lambda_r\bm{M}(\bm{k})\bm{q}_r.
  \label{eq:bloch-eigenproblem}
\end{equation}

Here $\bm k$ is the Bloch wave vector, and $\bm K(\bm k)$ and $\bm M(\bm k)$ are the Bloch-reduced stiffness and mass matrices; the stiffness dependence on $\alpha$ is suppressed in this equation. The vector $\bm q_r$ is the eigenvector of mode $r$, and $\lambda_r=\omega_r^2=(2\pi f_r)^2$, with angular eigenfrequency $\omega_r$ and eigenfrequency $f_r$. The fraction of modal mass in defect cell $X\in\{A,B\}$ is
\begin{equation}
  \eta_{X,r}
  =\frac{\bm{u}_r^{\mathrm{H}}\bm{M}_{X}^{\mathrm{phy}}\bm{u}_r}
        {\bm{u}_r^{\mathrm{H}}\bm{M}^{\mathrm{phy}}\bm{u}_r},
  \label{eq:mass-participation}
\end{equation}
where $\bm u_r$ is the full-space displacement reconstructed from $\bm q_r$ through the Bloch constraints, $\bm M_X^{\mathrm{phy}}$ is assembled only from elements in cell $X$, and $\bm M^{\mathrm{phy}}$ is the full physical mass matrix. The superscript $\mathrm H$ denotes conjugate transpose. Elementwise accumulation avoids ambiguity at shared and periodically identified nodes. Localization between cells $A$ and $B$ is measured by
\begin{equation}
  L_r=\frac{\eta_{A,r}-\eta_{B,r}}
            {\eta_{A,r}+\eta_{B,r}}.
  \label{eq:polarization}
\end{equation}

Thus $L_r>0$ favors $A$, $L_r<0$ favors $B$, and $L_r=0$ indicates equal modal mass in the two cells. A sign reversal along a continued branch changes the preferred defect. Here $|L_r|$ measures the relative imbalance between the two defect cells, while $\eta_{A,r}+\eta_{B,r}$ quantifies their combined share of the total modal mass. We count a sign change of $L_r$ at $\alpha\ne0$ as a substantial reversal if $|L_r|$ reaches at least 0.1 on both adjacent single-sign segments, each bounded by the next zero, the nominal state, or the end of the scan. For a harmonic mode, the common frequency factor cancels from the ratio of time-averaged kinetic energies, so these mass-weighted displacement fractions also describe the kinetic-energy distribution.

For softening we use $s=-\alpha$. On a branch that first favors $B$ and then favors $A$, $s_c>0$ denotes the sign-changing zero of $L$. The nominal model assigns one singlet and two doublet coordinates to each site. Its fixed basis includes the displacement of the surrounding host at the nominal state. We project the stiffness, the prescribed stiffness change in $B$, the mass, and the two physical cell-mass operators onto this basis once for each layout (Sec.~\ref{sec:nominal-operators}). To obtain the spectral reference $s_\times$, we use these same single-defect families. We suppress their mutual stiffness and mass blocks, preserve the site-pair blocks, and solve for the crossing of the specified pair of branches (Appendix~\ref{app:family-crossings}). Because each family subspace is fixed by single-defect symmetry, both $s_\times$ and the site bias of each crossing mode are invariant under nonsingular within-family coordinate transformations. All localization calculations use the same physical cell-mass operators, including when we compare different dynamical models.

\subsection{Branch Comparison and Reference Accuracy}
\label{sec:detuning-branch-tracking}

At an exact crossing, uncoupled channels retain their family identity; near an avoided crossing, coupled branches can exchange family character. Ordering modes by frequency alone would therefore compare different modal identities on the two sides of a crossing. To follow modal identity instead, modes of the reduced comparison models introduced in Sec.~\ref{sec:coupling-controls} and of the reference model are paired at the nominal state and tracked independently across discrete stiffness steps by maximizing the mass-weighted modal assurance criterion (MAC), the squared normalized mass inner product of two mode shapes~\citep{KimKim2000,Lu2020,Kuhn1955}. Nearly coincident frequencies are grouped into clusters using a relative frequency-gap threshold $g_{\mathrm{cl}}=10^{-4}$. Unresolved clusters are compared as subspaces and excluded from single-mode error statistics, whereas symmetry-isolated blocks resolve crossings directly (Sec.~\ref{app:tracking-cluster-metrics} of the supplementary material). At fixed detuning, matching each model directly to the reference state isolates which omitted interactions dominate the instantaneous error.

A parameter-enriched reduced-order model (ROM), solved directly in an uncondensed enriched subspace and verified against independent FOM solutions, provides the multimode reference for the PMMA--aluminum detuning sweeps in Sec.~\ref{sec:coupling-controls} (Secs.~\ref{app:family-retention} and~\ref{app:primary-rom-validation} of the supplementary material provide retained-family checks and reference validation). Its localization discrepancies from the FOM are negligible compared with the independent-channel errors examined below.

\section{Opposite Orders of Spectral Crossing and Spatial Reversal}
\label{sec:results}

\begin{figure}[!t]
  \centering
  \begin{minipage}[t]{0.49\linewidth}
    \centering\includegraphics[trim=0 6.5pt 0 0,clip,width=\linewidth,height=171.5pt,keepaspectratio]{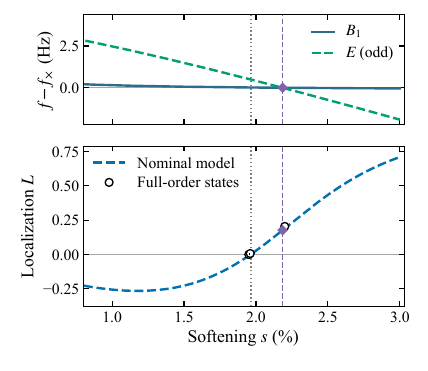}
    \par\smallskip{\small (a) PMMA--aluminum.\par}
  \end{minipage}\hfill
  \begin{minipage}[t]{0.49\linewidth}
    \centering\includegraphics[trim=0 6.5pt 0 0,clip,width=\linewidth,height=171.5pt,keepaspectratio]{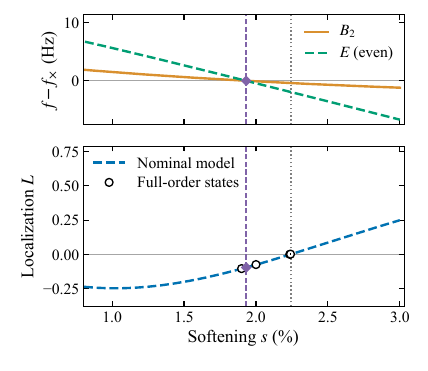}
    \par\smallskip{\small (b) Aluminum--tungsten.\par}
  \end{minipage}
  \caption{Uncoupled frequency offsets from $f_\times$ (top) and localization of the tracked coupled mode (bottom) in the two diagonal dimers. Positive $L$ favors $A$ and negative $L$ favors $B$. Purple diamonds mark the uncoupled crossing and dotted lines the nominal reversal. Open circles show independent full-order states. The black segment at $L=0$ spans their reversal bracket.}
  \label{fig:ordering}
\end{figure}

In Fig.~\ref{fig:ordering}, we follow the third nominal mode in the diagonal PMMA--aluminum dimer and the fourth in the aluminum--tungsten dimer [layout in Fig.~\ref{fig:numerical-geometry-schematic}(c)], counting in ascending frequency within the six retained modes. Both branches initially favor the softened defect $B$ and undergo one nonzero reversal toward $A$ over $0<s\leq0.03$. The relevant mirror sectors are odd in PMMA--aluminum and even in aluminum--tungsten. Each sector contains four coordinates: a singlet and one doublet partner at each site. In both dimers the upper singlet site-pair branch crosses the lower branch of the interacting doublet partner when their mutual coupling is removed.

In PMMA--aluminum the independent full-order solutions bracket reversal between 1.950\% and 1.960\% softening, so by the uncoupled crossing at 2.1850\% the coupled nominal mode already favors $A$ ($L=+0.1763$), consistent with the nearby full-order value $L=+0.2039$ at 2.200\%. In aluminum--tungsten the order is reversed: at the uncoupled crossing (1.9321\%), the coupled nominal mode still favors $B$ ($L=-0.0966$), as do the bracketing full-order solutions ($L=-0.1048$ at 1.900\% and $-0.0753$ at 2.000\%), and the full-order branch changes sign later within 2.235\%--2.240\% softening.

\section{Site Preference from the Two Uncoupled Crossing Modes}
\label{sec:criterion}

Let $\bm u_b$ and $\bm u_e$ be the real singlet and doublet-partner eigenvectors at $s_\times$, each normalized to unit modal mass in its family block. Write $\bm V=[\bm u_b,\bm u_e]$ in the common nominal coordinates. Their uncoupled squared angular frequencies coincide at $\lambda_\times$. Restoring the coupling gives the projected stiffness and mass
\begin{equation}
 \bm K_\times=\begin{pmatrix}\lambda_\times&k\\k&\lambda_\times\end{pmatrix},
 \qquad
 \bm M_\times=\begin{pmatrix}1&m\\m&1\end{pmatrix},
 \label{eq:crossing-pair}
\end{equation}
where $k$ and $m$ are the cross-family matrix elements, with $|m|<1$ by positive definiteness of the mass matrix. When $k-\lambda_\times m\ne0$, the two eigenvalues $(\lambda_\times+\sigma k)/(1+\sigma m)$ ($\sigma=\pm1$) are distinct, and their mass-normalized eigenvectors are $(1,\sigma)^{\mathrm{T}}/\sqrt{2(1+\sigma m)}$. The two coefficients therefore have equal magnitude at the crossing even when the basis modes have nonzero mass overlap.

Define $\bm N=\bm Q_A-\bm Q_B$ and $\bm Q=\bm Q_A+\bm Q_B$ from the physical cell-mass operators projected into the common basis. The individual site biases $d_b=\bm u_b^{\mathrm{T}}\bm N\bm u_b$ and $d_e=\bm u_e^{\mathrm{T}}\bm N\bm u_e$ are signed fractions of total modal mass. The cross-mode matrix element $d_{be}=\bm u_b^{\mathrm{T}}\bm N\bm u_e$ enters the numerator as the interference term $2\sigma d_{be}$. For positive defect-cell mass, the sign of $L$ is the sign of its numerator. In the two-state model at $s_\times$ this gives
\begin{equation}
 D_\sigma=\frac{d_b+d_e+2\sigma d_{be}}{2(1+\sigma m)},\qquad
 |d_b+d_e|>2|d_{be}|\quad\Longrightarrow\quad
 \operatorname{sgn}L_\sigma=\operatorname{sgn}(d_b+d_e).
 \label{eq:crossing-bias}
\end{equation}

Here $\sigma$ specifies the relative sign of the two coefficients. Reversing the sign convention for a basis mode exchanges the labels and preserves the two physical predictions. Although the two coupled states are mass-orthogonal, both can concentrate preferentially at the same site.

\begin{table}[!b]
\centering
\caption{Direction criterion at the uncoupled crossings. The first four rows use only the two uncoupled modes. The last two rows compare the target state of their coupled projection with that of the complete nominal mirror sector.}
\label{tab:crossing-bias}
\begin{tabular}{lrr}
\toprule
Quantity & PMMA--aluminum & Aluminum--tungsten\\
\midrule
Singlet bias $d_b$ & $+0.70381$ & $+0.32745$\\
Doublet-partner bias $d_e$ & $-0.50161$ & $-0.72975$\\
$|d_{be}|$ & $0.000562$ & $0.000314$\\
$2|d_{be}|/|d_b+d_e|$ & 0.556\% & 0.156\%\\
Two-state $L(s_\times)$ & $+0.1422$ & $-0.2840$\\
Complete nominal $L(s_\times)$ & $+0.1763$ & $-0.0966$\\
\bottomrule
\end{tabular}
\end{table}

In both pairs the singlet favors $A$ and the doublet partner favors $B$ (Table~\ref{tab:crossing-bias}), but the singlet bias is stronger in PMMA--aluminum whereas the doublet-partner bias is stronger in aluminum--tungsten, yielding opposite net biases. Because the interference term is too small in each case to alter the sign of either two-state combination, both coupled states at the crossing favor the same site. The target branches of the complete sectors reproduce these signs, and their companion modes, paired at the crossing by mass overlap, likewise favor the predicted sites ($L=+0.0963$ and $-0.3125$, respectively). Given the single $B$-to-$A$ sign change on each target branch in Fig.~\ref{fig:ordering}, reversal occurs before the crossing when the target already favors $A$, and after it when the target still favors $B$.

\section{Multimode Mass Balance and Prediction of the Reversal Threshold}
\label{sec:nominal-threshold}

\subsection{Balance Between Families on the Coupled Branch}

The crossing pair gives the sign of the spatial preference in both dimers. To locate the sign change, we follow the mass-normalized eigenvector $\bm y(s)$ of the complete nominal model. For each family $g=b,e$ define its contributions $n_g$ and $q_g$ to the difference and sum of the two defect-cell modal masses from the respective diagonal blocks of $\bm N$ and $\bm Q$. The family bias is $\ell_g=n_g/q_g$, and the singlet share of the summed family-block defect mass is $p=q_b/(q_b+q_e)$. With $n_x$ the cross-family contribution to the difference and $\xi=n_x/(q_b+q_e)$, the full numerator is
\begin{equation}
 D=n_b+n_e+n_x=(q_b+q_e)(\ell_b-\ell_e)(p-p_*),
 \qquad p_*=\frac{-\ell_e-\xi}{\ell_b-\ell_e}.
 \label{eq:body-family-balance}
\end{equation}

This form applies where $q_b,q_e>0$ and $\ell_b\ne\ell_e$ (Appendix~\ref{app:family-mass-balance}). The single-mode biases $d_b,d_e$ in Eq.~\eqref{eq:crossing-bias} describe uncoupled states. In Eq.~\eqref{eq:body-family-balance} both $p$ and the share $p_*$ needed to balance the two sites are evaluated on the evolving coupled mode.

\begin{figure}[!t]
  \centering
  \begin{minipage}[t]{0.49\linewidth}
    \centering\includegraphics[trim=0 6.5pt 0 0,clip,width=\linewidth,height=125.5pt,keepaspectratio]{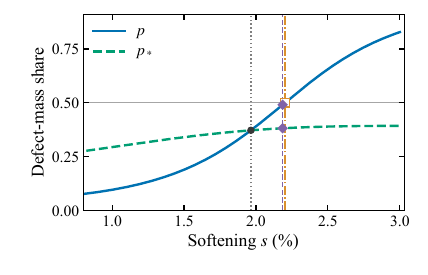}
    \par\smallskip{\small (a) PMMA--aluminum.\par}
  \end{minipage}\hfill
  \begin{minipage}[t]{0.49\linewidth}
    \centering\includegraphics[trim=0 6.5pt 0 0,clip,width=\linewidth,height=125.5pt,keepaspectratio]{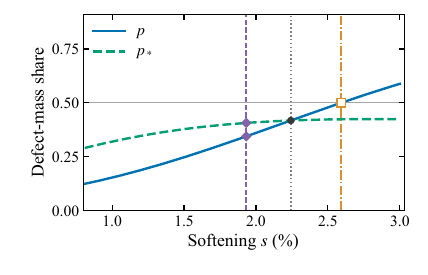}
    \par\smallskip{\small (b) Aluminum--tungsten.\par}
  \end{minipage}
  \caption{Singlet share $p$ and the share $p_*$ required for equal defect-cell modal kinetic energies along the two nominal branches. The curves intersect at the spatial reversal. Diamonds mark the uncoupled crossing, the filled circle marks reversal, and the square marks equal family shares ($p=1/2$).}
  \label{fig:diagonal-balance}
\end{figure}

Figure~\ref{fig:diagonal-balance} places both dimers on the same softening scale. At the PMMA--aluminum crossing the singlet supplies 49.09\% of the summed family-block defect mass, already above the 38.19\% needed for balance at that state. At the aluminum--tungsten crossing it supplies 34.40\%, still below the required 40.67\%. The respective cross-family contributions $n_x$ at the crossings are $+0.000536$ and $+0.000328$ in units of total modal mass. The PMMA--aluminum branch reaches balance at a singlet share of 37.23\% and reaches equal family shares later, at 2.2014\% softening. Although the aluminum--tungsten branch reverses after its spectral crossing, it likewise reaches spatial balance before equal family shares, with the singlet supplying 41.79\% at reversal and reaching 50\% at 2.5916\% softening.

\subsection{Additional Modes of the Same Mirror Symmetry}

Keeping only the two crossing modes as a fixed basis gives reversal at 2.0287\% in PMMA--aluminum and 2.8856\% in aluminum--tungsten. Restoring the other site-pair modes of the same mirror symmetry gives 1.9648\% and 2.2425\%, respectively. In aluminum--tungsten the two-state approximation overestimates the distance from crossing to reversal by a factor of $3.07$ relative to the complete sector.

At the uncoupled crossing, the mass-orthogonal projection of the complete target state onto the crossing pair retains 99.28\% of the modal mass in PMMA--aluminum and 91.37\% in aluminum--tungsten (Appendix~\ref{app:pair-projection}). In aluminum--tungsten, the two-state eigenvector gives $L=-0.2840$, the projection of the complete eigenvector gives $L=-0.4566$, and the complete state gives $L=-0.0966$. The corresponding contributions to $D$ are $-0.3072$ from the projected field, $+0.2091$ from interference with the complement, and $+0.0272$ from the complement alone.

At the aluminum--tungsten nominal reversal ($s_c=2.2425\%$), decompose the mass-normalized state as $\bm y=\bm y_\parallel+\bm y_\perp$, where $\bm y_\parallel$ is its mass-orthogonal projection onto the crossing pair frozen at $s_\times$ (Appendix~\ref{app:pair-projection}) and $\bm y_\perp=\bm y-\bm y_\parallel$ is the complementary displacement. This projected component carries 92.17\% of the modal mass and still favors the softened defect ($L=-0.3461$). The three contributions to the mass difference, proportional to the kinetic-energy difference at this frequency, are
\[
\begin{aligned}
 D &= \bm y_\parallel^{\mathrm{T}}\bm N\bm y_\parallel
      +2\bm y_\parallel^{\mathrm{T}}\bm N\bm y_\perp
      +\bm y_\perp^{\mathrm{T}}\bm N\bm y_\perp \\
   &\simeq -0.23247+0.21247+0.02000=0.
\end{aligned}
\]

Interference with the remaining same-symmetry displacement thus supplies most of the compensating bias toward $A$. Because both $\bm y_\parallel$ and $\bm y_\perp$ contain singlet and doublet components, their interference includes within-family contributions, unlike $n_x$ in Eq.~\eqref{eq:body-family-balance}, which contains only singlet--doublet cross terms. Even at 3\% softening in aluminum--tungsten, the mass-orthogonal projection onto the frozen crossing pair still favors $B$ ($L=-0.0933$), whereas the complete nominal and full-order modes both favor $A$ ($L=+0.2492$ and $+0.2531$, respectively).

\subsection{Prediction from Nominal Operators}
\label{sec:nominal-operators}

For a given layout, let $\bm V_r$ contain the $r$ physical displacement fields retained in a fixed basis, including their host response. Projection gives
\begin{equation}
\begin{aligned}
 \bm K_r^0 &= \bm V_r^{\mathrm H}\bm K_0^{\mathrm{phy}}\bm V_r, &
 \bm K_{rB} &= \bm V_r^{\mathrm H}\bm K_B^{\mathrm{phy}}\bm V_r,\\
 \bm M_r &= \bm V_r^{\mathrm H}\bm M^{\mathrm{phy}}\bm V_r, &
 \bm Q_{Xr} &= \bm V_r^{\mathrm H}\bm M_X^{\mathrm{phy}}\bm V_r,\quad X=A,B.
\end{aligned}
\label{eq:nominal-projected-operators}
\end{equation}

With $\bm N_r=\bm Q_{Ar}-\bm Q_{Br}$ and $\bm Q_r=\bm Q_{Ar}+\bm Q_{Br}$, the eigenproblem and localization along the continued branch are
\begin{equation}
 (\bm K_r^0-s\bm K_{rB})\bm y=\lambda\bm M_r\bm y,\qquad
 L(s)=\frac{\bm y^{\mathrm H}\bm N_r\bm y}{\bm y^{\mathrm H}\bm Q_r\bm y}.
 \label{eq:nominal-projected-pencil}
\end{equation}

The first $B$-to-$A$ sign change of the numerator at $s>0$, with positive defect mass in the denominator, gives $s_c$. The softening magnitude $s$ scales the prescribed defect stiffness $\bm K_{rB}$. All operators are obtained from the nominal structure, without fitting detuned branches (see Table~\ref{tab:case-specific-rom-parameters} of the supplementary material for the recovery bases).

In each diagonal dimer, mirror symmetry confines the target branch of the six-coordinate nominal model to four coordinates, one singlet and one doublet partner at each site. Solving in this invariant sector gives the complete nominal threshold. Projecting further onto the two modes at $s_\times$ gives the frozen-pair predictions in Table~\ref{tab:threshold-accuracy}. The two diagonal nominal thresholds lie just above the upper ends of the full-order brackets, with relative errors of 0.65\% and 0.26\% against the linearly interpolated roots in PMMA--aluminum and aluminum--tungsten, respectively.

\begin{table}[!b]
\centering
\caption{Softening at localization reversal, expressed as a percentage. The complete nominal model retains both doublet partners and the singlet at each site. The two-state basis is frozen at the nominal uncoupled crossing, which also serves as the common reference for the diagonal offsets reported below.}
\label{tab:threshold-accuracy}
\setlength{\tabcolsep}{4pt}
\begin{tabular}{lccc}
\toprule
System and layout & Frozen pair & Complete nominal & Full-order bracket\\
\midrule
PMMA--aluminum, diagonal & 2.0287 & 1.9648 & 1.950--1.960\\
Aluminum--tungsten, diagonal & 2.8856 & 2.2425 & 2.235--2.240\\
PMMA--aluminum, off-axis & --- & 2.1229 & 2.100--2.110\\
\bottomrule
\end{tabular}
\end{table}

The signed offset $\delta s=s_c-s_\times$ measures how far reversal precedes or follows the uncoupled crossing. Using the same nominal uncoupled crossing for both estimates in each dimer, the predicted and full-order interpolated offsets are $-0.2202$ and $-0.2328$ softening percentage points in PMMA--aluminum, and $+0.3104$ and $+0.3045$ in aluminum--tungsten. Relative to the magnitudes of the interpolated offsets, the errors are 5.42\% and 1.94\%. Both full-order samples bracketing each reversal lie on the same side of its respective crossing (Fig.~\ref{fig:ordering}), so the event ordering follows directly from the sampled states.

Off axis, both doublet partners can mix with the singlet. Retaining only $B_1$ and the second nominal doublet channel $E_2$ (defined in Sec.~\ref{sec:comparison-models}) at the two sites gives $s_c=2.1703\%$ in PMMA--aluminum. Restoring the other doublet channel improves the prediction to 2.1229\% (Table~\ref{tab:threshold-accuracy}). Figure~\ref{fig:nominal-threshold} in Appendix~\ref{app:nominal-threshold-derivation} compares the truncated branch with full-order results near reversal. A $7\times7$ supercell at the same offset $(2,1)$ also reverses on the tracked branch, at a larger softening (Sec.~\ref{app:primary-rom-validation} of the supplementary material).

\section{Coupling Controls and Layout Dependence}
\label{sec:coupling-controls}

The diagonal results above rely on cross-family coupling within a single mirror sector. Next, we test controlled truncations across the axial, off-axis, and diagonal layouts of the PMMA--aluminum system [Fig.~\ref{fig:numerical-geometry-schematic}(a)--(c)] to separate dynamical cross-family coupling from both cross-family observation terms and intra-doublet coupling, and to examine how dimer symmetry restricts which layouts exhibit reversal.

\subsection{Comparison Models}
\label{sec:comparison-models}
For the PMMA--aluminum dimer, a channel is one fixed direction in a single-defect family with one coordinate at each site. The three comparison models use the same ordered coordinates,
\begin{equation}
\bm x=[x_{B_1A},x_{B_1B},x_{E_1A},x_{E_1B},x_{E_2A},x_{E_2B}]^{\mathrm T},
\label{eq:fixed-channel-coordinates}
\end{equation}
and share the same host recovery and physical defect-cell mass operators. Table~\ref{tab:coupling-levels} identifies the blocks retained in their condensed stiffness and mass matrices. The nominal doublet channels are held fixed during detuning (Secs.~\ref{app:fixed-channel-construction} and~\ref{app:physical-reconstruction} of the supplementary material).

\begin{table}[!b]
 \centering
 \caption{Coupling levels of the condensed models. Within-channel coupling between defects $A$ and $B$ is retained at every level. Deleted blocks are removed from both stiffness and mass matrices.}
 \label{tab:coupling-levels}
 \setlength{\tabcolsep}{6pt}
 \begin{tabular}{lcc}
  \toprule
  Model & Cross-family coupling & Coupling between $E$ channels\\
  \midrule
  Fully coupled ($k=0$) & Retained & Retained\\
  Family-separated ($k=1$) & Removed & Retained\\
  Independent-channel ($k=2$) & Removed & Removed\\
  \bottomrule
 \end{tabular}
\end{table}

\subsection{Localization Reversal Along a Continued Branch}
\label{sec:primary-detuning-response}

\begin{figure}[!t]
  \centering
  \begin{minipage}[t]{0.45\linewidth}
    \centering
    \includegraphics[width=\linewidth,height=120pt,keepaspectratio]{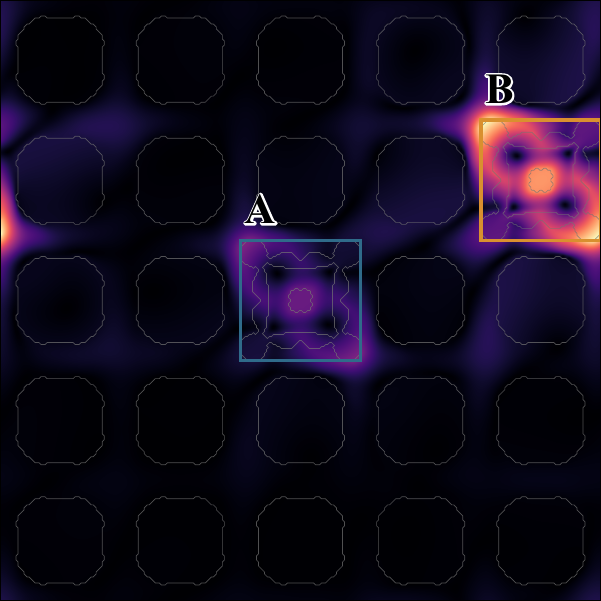}
    \par\smallskip{\small (a) Independent-channel model: \SI{1197.434}{Hz}, $L=-0.751$.\par}
  \end{minipage}
  \hfill
  \begin{minipage}[t]{0.45\linewidth}
    \centering
    \includegraphics[width=\linewidth,height=120pt,keepaspectratio]{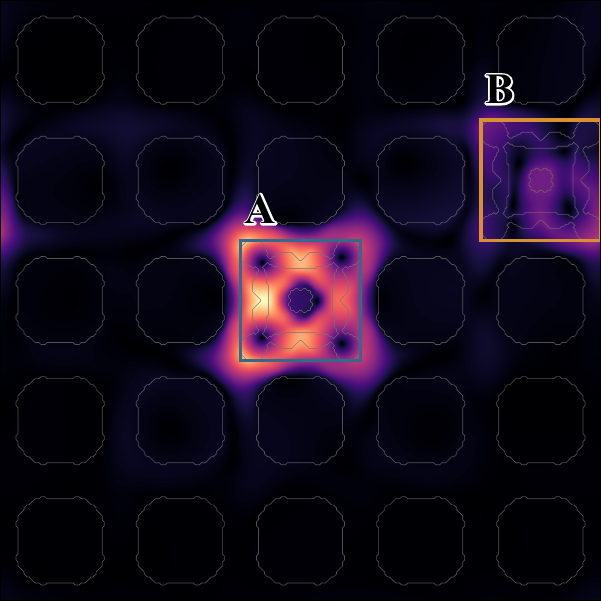}
    \par\smallskip{\small (b) FOM: \SI{1199.307}{Hz}, $L=+0.746$.\par}
  \end{minipage}
  \par\medskip
  \includegraphics[width=190bp]{transfer_displacement_colorbar.pdf}
  \par\medskip
  \begin{minipage}[t]{0.48\linewidth}
    \centering
    \includegraphics[trim=0 6.5pt 0 0,clip,width=\linewidth,height=139pt,keepaspectratio]{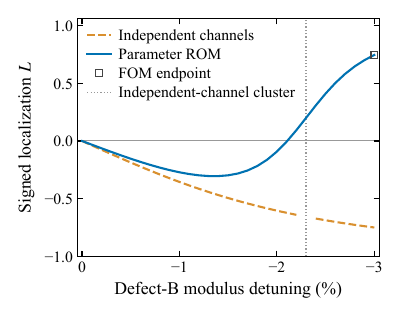}
    \par\smallskip{\small (c) Localization along independently continued branches.\par}
  \end{minipage}
  \hfill
  \begin{minipage}[t]{0.48\linewidth}
    \centering
    \includegraphics[trim=0 6.5pt 0 0,clip,width=\linewidth,height=139pt,keepaspectratio]{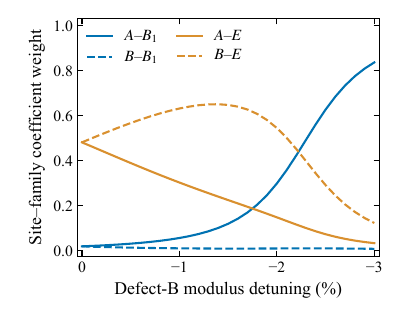}
    \par\smallskip{\small (d) Parameter-ROM coefficient weights.\par}
  \end{minipage}
\caption{Off-axis localization. (a,b) Independently normalized endpoint fields at $\alpha=-0.03$. The FOM mode is matched to the reference ROM by physical MAC. (c) Independently continued branches. Clustering excludes the independent-channel comparison at $\alpha=-0.023$. (d) Normalized coefficient weights from Eq.~\eqref{eq:site-family-coefficients} of the supplementary material.}
  \label{fig:offaxis-wrong-defect}
\end{figure}

Figure~\ref{fig:offaxis-wrong-defect} follows the lower-frequency nominal mode of the off-axis $E_2$ channel and its matched reference branch. Both initially favor the softened defect $B$. As softening continues, the reference mode shifts toward $A$, where its $B_1$ coefficient weight reaches 0.836 at $\alpha=-0.03$, consistent with family exchange near curve veering~\citep{PerkinsMote1986,Pierre1988Localization}. At that same detuning the independent-channel and full-order modes give $L=-0.751$ and $+0.746$, respectively. Their frequency difference is only 0.156\% of the full-order frequency, but amounts to 84\% of the spacing to the nearest retained full-order mode, a scale relevant to eigenvector sensitivity~\citep{DavisKahan1970}.

\subsection{Dynamical Coupling and Localization Measures at Fixed Mode Shapes}
\label{sec:fixed-branch-results}

Cross-family interactions can affect localization in two ways: coupling blocks in the stiffness and mass matrices alter the eigenvector, while cross-family terms in the physical mass quotient alter the localization value assigned to a given state. We separate these effects by first deleting dynamical coupling and solving the modified eigenproblem, and then keeping the coupled eigenvector fixed while removing cross-family terms only from the localization measure.

\begin{figure}[!t]
  \centering
  \begin{minipage}[t]{0.48\linewidth}
    \centering
    \includegraphics[trim=0 5.5pt 0 0,clip,width=\linewidth,height=139.5pt,keepaspectratio]{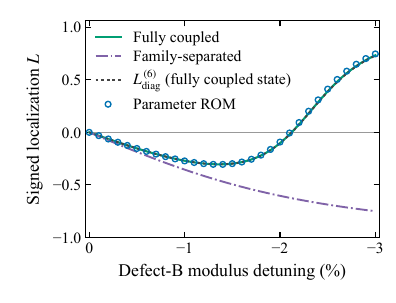}
    \par\smallskip{\small (a) Off-axis localization.\par}
  \end{minipage}
  \hfill
  \begin{minipage}[t]{0.48\linewidth}
    \centering
    \includegraphics[trim=0 5.5pt 0 0,clip,width=\linewidth,height=139.5pt,keepaspectratio]{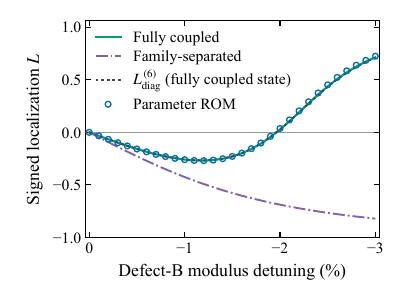}
    \par\smallskip{\small (b) Diagonal localization.\par}
  \end{minipage}
  \par\medskip
  \begin{minipage}[t]{0.48\linewidth}
    \centering
    \includegraphics[trim=0 5.5pt 0 0,clip,width=\linewidth,height=99.5pt,keepaspectratio]{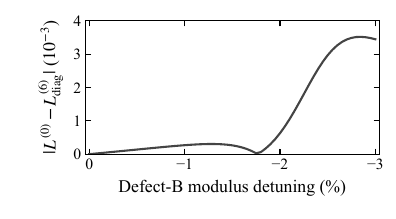}
    \par\smallskip{\small (c) Off-axis difference in localization index.\par}
  \end{minipage}
  \hfill
  \begin{minipage}[t]{0.48\linewidth}
    \centering
    \includegraphics[trim=0 5.5pt 0 0,clip,width=\linewidth,height=99.5pt,keepaspectratio]{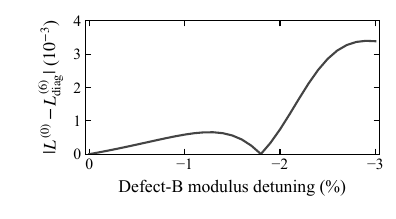}
    \par\smallskip{\small (d) Diagonal difference in localization index.\par}
  \end{minipage}
\caption{Effect of cross-family coupling on localization. (a,b) Independently continued fully coupled and family-separated branches. (c,d) Effect of omitting cross-family terms from the localization measure at fixed fully coupled mode shape. The two condensed models use 61 off-axis and 31 diagonal samples, and the reference ROM uses 31 in each layout.}
  \label{fig:fixed-branch-continuation}
\end{figure}

Figure~\ref{fig:fixed-branch-continuation} compares the fully coupled and family-separated models, both retaining the two doublet channels, along independently continued branches over $\alpha\in[-0.03,0]$ against the parameter-ROM reference. The off-axis target is the $E_2$ branch introduced above, and the diagonal PMMA--aluminum target is initialized from the lower-frequency nominal $E_1$ channel mode; coupled initial states are selected by maximum overlap with these targets.

A separate comparison of localization measures holds the fully coupled eigenvector $\bm{x}^{(0)}\in\mathbb{C}^6$ fixed. Let $\bm{N}_c=\bm{Q}_A^c-\bm{Q}_B^c$ and $\bm{Q}_c=\bm{Q}_A^c+\bm{Q}_B^c$, where $\bm Q_X^c(\alpha)=\bm J(\alpha)^{\mathrm H}\bm M_X^{\mathrm{phy}}\bm J(\alpha)$ is the recovered modal-mass form over defect cell $X\in\{A,B\}$.

For single-defect modes centered on the same cell, $C_{4v}$ symmetry makes the cell-mass inner product between $B_1$ and $E$ vanish. Because the recovered forms $\bm Q_X^c$ also include the partner defect's evanescent tail and the host response, the lower dimer symmetry permits nonzero $B_1$--$E$ cross terms. Setting only those $B_1$--$E$ blocks of $\bm N_c$ and $\bm Q_c$ to zero defines $\bm N_c^{\mathrm{diag}}$ and $\bm Q_c^{\mathrm{diag}}$, and gives the localization measure with block-diagonal mass forms
\begin{equation}
  L_{\mathrm{diag}}^{(6)}(\bm{x}^{(0)})=
  \frac{(\bm{x}^{(0)})^{\mathrm{H}}\bm N_c^{\mathrm{diag}}\bm{x}^{(0)}}
       {(\bm{x}^{(0)})^{\mathrm{H}}\bm Q_c^{\mathrm{diag}}\bm{x}^{(0)}}.
  \label{eq:fixed-state-observable}
\end{equation}

Here the subscript $\mathrm{diag}$ denotes block diagonality between the $B_1$ and $E$ families in the six-coordinate space, retaining all $E_1$--$E_2$ and inter-site terms within each family. Both $L$ and $L_{\mathrm{diag}}^{(6)}$ are evaluated directly on the same coupled state $\bm{x}^{(0)}$.

The fully coupled condensed model reproduces the off-axis and diagonal reversals, with maximum discrepancies in $L$ from the reference ROM below 0.022 on common samples of these continued branches. The family-separated model instead favors defect $B$ at every nonzero sample in both layouts: deleting the cross-family stiffness and mass blocks eliminates reversal even though both doublet channels and their mutual coupling remain intact. In the diagonal layout, reflection symmetry isolates the $(B_1,E_1)$ block from $E_2$ (where the nominal channels $E_1$ and $E_2$ coincide with the mirror-odd and mirror-even diagonal partners $E_\perp$ and $E_\parallel$ of Table~\ref{tab:dimer-small-groups}), resolving crossings with the complementary block (see Sec.~\ref{app:sixdim-controls} of the supplementary material for endpoint values and a target-only coupling control).

When the coupled mode shapes are held fixed, omitting the cross-family terms in the localization measure changes $L$ by less than 0.0036 and preserves both reversal intervals. Because deleting the cross-family dynamical blocks removes reversal whereas omitting the cross-family terms in the localization measure leaves it intact, dynamical redistribution between modal families is the dominant mechanism.

In aluminum--tungsten, deleting the cross-family stiffness and mass blocks keeps the tracked $E$ branch localized on $B$, with $L=-0.979$ at 3\% softening. With the coupled mode shape held fixed, omitting the cross-family observation term alone shifts the nominal threshold from 2.2425\% to 2.2439\%. Both operations use the same partition into singlet and doublet families.

\subsection{Symmetry and Layout Dependence}
\label{sec:primary-dimer-small-groups}

Table~\ref{tab:dimer-small-groups} gives the symmetry sectors before and after detuning. Here $E_+$ and $E_-$ denote the even and odd axial channels, and $E_{\parallel}$ and $E_{\perp}$ the even and odd diagonal channels. Only modes in the same mirror sector can mix when the reflection is preserved. The full induced representations are derived in Appendix~\ref{app:coordinate-covariance}.

\begin{table}[!b]
  \centering
\caption{Symmetry-allowed cross-family mixing in the PMMA--aluminum dimer.}
  \label{tab:dimer-small-groups}
  \setlength{\tabcolsep}{3pt}
  \begin{tabular}{L{0.11\linewidth}C{0.08\linewidth}L{0.30\linewidth}C{0.12\linewidth}L{0.30\linewidth}}
    \toprule
    Layout & $G(0)$ & Allowed mixing at $\alpha=0$ & $G(\alpha\ne0)$ & Effect of detuning \\
    \midrule
    Axial & $C_{2v}$ & $B_1,E_+$: two shared sectors; $E_-$ separate & $C_s$ & $B_1,E_+$: same reflection sector \\
    Off-axis & $C_2$ & $B_1,E_+,E_-$: together in both sectors & $C_1$ & All six coordinates in one sector \\
    Diagonal & $C_{2v}$ & $B_1,E_{\perp}$: two shared sectors; $E_{\parallel}$ separate & $C_s$ & $B_1,E_{\perp}$: same reflection sector \\
    \bottomrule
  \end{tabular}
\end{table}

Mirror symmetry in the axial and diagonal layouts excludes one $E$ partner from mixing with $B_1$, whereas both $E$ partners can mix with $B_1$ off axis; these selection rules hold across all detunings. Across the PMMA--aluminum branches, the nominal model and reference ROM both give six substantial reversals off axis, four diagonally, and none axially (Table~\ref{tab:primary-layout-roots}). Thus the axial and diagonal layouts admit a singlet--doublet coupling but differ in the resulting redistribution.

A pointwise restoration comparison at matched reference states further shows that omitted $B_1$--$E$ coupling dominates the large localization errors, exceeding each separately evaluated condensation, subspace-restriction, and enrichment correction (Sec.~\ref{app:pointwise-restoration} of the supplementary material).

\subsection{Effect of Cross-Family Coupling Strength}

\begin{figure}[!b]
 \centering
 \begin{minipage}[t]{0.48\linewidth}
   \centering
   \includegraphics[trim=0 7.5pt 0 0,clip,width=\linewidth,height=147.5pt,keepaspectratio]{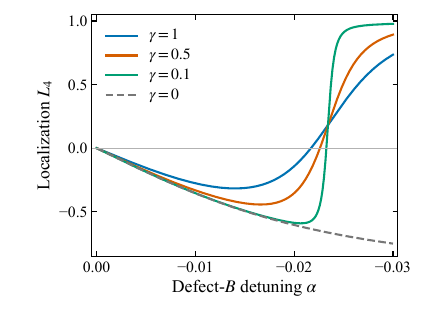}
   \par\smallskip{\small (a) Localization under coupling variation.\par}
 \end{minipage}
 \hfill
 \begin{minipage}[t]{0.48\linewidth}
   \centering
   \includegraphics[trim=0 7.5pt 0 0,clip,width=\linewidth,height=147.5pt,keepaspectratio]{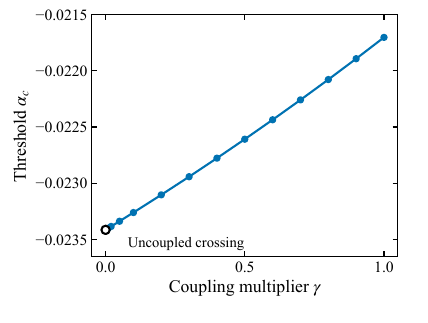}
   \par\smallskip{\small (b) Threshold movement.\par}
 \end{minipage}
 \caption{Coupling-strength dependence in the nominal four-coordinate model. At $\gamma=0$, continuation preserves the $E_2$ family. The open symbol is an uncoupled frequency crossing, not a localization zero.}
 \label{fig:coupling-strength}
\end{figure}

Figure~\ref{fig:coupling-strength} shows the effect of varying the cross-family coupling in the off-axis four-coordinate model. Writing $\mathcal D$ for the operator that retains the two diagonal family blocks ($B_1,E_2$), we set
\begin{equation}
 \begin{gathered}
 \bm X_\gamma=\mathcal D(\bm X)+\gamma[\bm X-\mathcal D(\bm X)],\\
 \bm X\in\{\bm K_4^0,\bm K_{4B},\bm M_4\},\qquad 0\leq\gamma\leq1.
 \end{gathered}
 \label{eq:coupling-strength-intervention}
\end{equation}

The nominal recovery and physical forms $\bm N_4,\bm Q_4$ remain fixed. The dimensionless multiplier $\gamma$ scales the cross-family blocks of the stiffness and generalized mass matrices at fixed geometry. The nominal target is continued from $\gamma=1$ along $\alpha=0$, then along detuning at each $\gamma$, with overlaps measured in the original physical mass metric.

For every positive multiplier tested, the continued branch reverses. As coupling weakens, reversal approaches the uncoupled crossing and becomes sharper (Fig.~\ref{fig:coupling-strength} and Appendix~\ref{app:nominal-threshold-derivation}). At $\gamma=0$ the families are invariant: the continued $E_2$ branch passes through the crossing and still favors $B$ at 3\% softening ($L_4=-0.7505$). Switching branches by frequency rank there would change the family being followed.

\section{Conclusions}
\label{sec:conclusions}

Under progressive softening of one defect, a localized mode switches to the unchanged defect before the uncoupled family crossing in the diagonal PMMA--aluminum dimer and after it in the aluminum--tungsten dimer. The sign of the summed site biases of the two uncoupled modes gives the preferred defect at each crossing because the interference between these modes is small. The singlet bias toward the unchanged defect dominates in PMMA--aluminum, whereas the doublet-partner bias toward the softened defect dominates in aluminum--tungsten. Since each tracked branch reverses once, these opposite preferences account for the opposite event orderings.

The reversal threshold depends on how the two defect cells reach equal kinetic energy along the coupled branch. They reach balance at singlet shares of 37.23\% and 41.79\% of the summed family-block defect mass in PMMA--aluminum and aluminum--tungsten, respectively. In aluminum--tungsten, the displacement projected onto the crossing pair still favors the softened site at the nominal reversal. Its interference with the rest of the same-symmetry displacement accounts for most of the compensating bias toward the unchanged site. With these components included, the nominal model places both thresholds within 1\% of the independently interpolated full-order values.

\section*{Supplementary Material}
The supplementary material contains the numerical settings and reference validation, the construction and physical recovery of the reduced models, and the continuation, cluster-comparison, and error-attribution procedures supporting the main results.

\section*{Acknowledgments}
This work was supported by MEXT KAKENHI Grant Number 25H00756 and The Kajima Foundation.

\section*{Data Availability}
The data that support the findings of this study are available within the article and its supplementary material, and additional numerical data are available from the corresponding author upon reasonable request.

\appendix
\counterwithin{equation}{section}
\renewcommand{\theequation}{\Alph{section}\arabic{equation}}
\renewcommand{\thetable}{\arabic{table}}

\section{Dimer Symmetry and Allowed Coupling}
\label{app:coordinate-covariance}

The local basis, whitening maps, and interaction operators used below are defined in Sec.~\ref{app:scalar-applicability-construction} of the supplementary material; the physical reconstruction is given in Sec.~\ref{app:physical-reconstruction} of the supplementary material.

Local reparameterization followed by rewhitening preserves coupling singular values. Family-resolved quantities are invariant under transformations preserving the labeled $B_1\oplus E$ decomposition.

For a symmetry operation $g\in G(\alpha)$ with physical displacement transformation $\bm T_g$ and the reconstructed mass-orthonormal basis $\bm\Phi^{\mathrm{phy}}$, the restricted symmetry representation is $\bm R_g=(\bm\Phi^{\mathrm{phy}})^{\mathrm H}\bm M^{\mathrm{phy}}\bm T_g\bm\Phi^{\mathrm{phy}}$, with character $\chi_g=\operatorname{tr}\bm R_g$, where $\operatorname{tr}$ denotes the matrix trace. In the local PMMA--aluminum order, the twofold rotation and axial reflections have representations
\begin{equation}
  \begin{aligned}
  \bm{D}_{\mathrm{loc}}(C_2)&=\operatorname{diag}(1,-1,-1),\\
  \bm{D}_{\mathrm{loc}}(\sigma_x)&=\operatorname{diag}(1,1,-1),\\
  \bm{D}_{\mathrm{loc}}(\sigma_y)&=\operatorname{diag}(1,-1,1),\\
  \bm{D}^{(0)}_6(g)&=\bm{\Pi}(g)\otimes\bm{D}_{\mathrm{loc}}(g).
  \end{aligned}
  \label{eq:raw-dimer-representation}
\end{equation}

Here $\bm\Pi(g)$ permutes sites, and $\otimes$ denotes the Kronecker product. Dimer operations are defined about the midpoint, with unit Bloch phase for accompanying supercell translations at $\Gamma$. Their whitened representations are $\bm D_\mu(g)=\bm T_{\mathrm{loc}}^{-1}\bm D_6^{(0)}(g)\bm T_{\mathrm{loc}}$, where $\bm T_{\mathrm{loc}}$ is the local mass-whitening map. For the remaining group $G(\alpha)$, the projector onto irrep $\zeta$ is
\begin{equation}
  \bm{P}_{\zeta}^{\mu}
  =\frac{d_{\zeta}}{|G(\alpha)|}
  \sum_{g\in G(\alpha)}\chi_{\zeta}^{*}(g)\bm{D}_{\mu}(g),
  \label{eq:dimer-group-projector}
\end{equation}
where $\zeta$ labels an irrep, $d_\zeta$, $\chi_\zeta$, and $|G(\alpha)|$ denote its dimension, character, and the group order; $*$ denotes complex conjugation. Covariance of the whitened overlap mass $\bm S_\mu$ and interaction operator $\bm C_\mu(\alpha)$ requires $\bm D_\mu^{\mathrm H}\bm S_\mu\bm D_\mu=\bm S_\mu$ and $\bm D_\mu^{\mathrm H}\bm C_\mu\bm D_\mu=\bm C_\mu$, giving
\begin{equation}
  (\bm{P}_{\zeta}^{\mu})^{\mathrm{H}}
  \bm{C}_{\mu}(\alpha)\bm{P}_{\xi}^{\mu}=\bm{0},
  \qquad \zeta\ne\xi.
  \label{eq:dimer-selection-rule}
\end{equation}

\begin{table}[!t]
  \centering
  \caption{Induced dimer representations in PMMA--aluminum locally whitened coordinates. Entries in braces share a symmetry sector and may mix.}
  \label{tab:full-dimer-decomposition}
  \setlength{\tabcolsep}{4pt}
  \begin{tabular}{L{0.16\linewidth}C{0.10\linewidth}L{0.22\linewidth}L{0.42\linewidth}}
    \toprule
    Layout and parameter & Group & Decomposition & Symmetry sectors in single-defect-family coordinates \\
    \midrule
    Axial, $\alpha=0$ & $C_{2v}$ & $2A_1^{(d)}+A_2^{(d)}$\newline $+2B_x^{(d)}+B_y^{(d)}$ & $A_1^{(d)}:\{B_1^+,E_+^-\}$;\newline $B_x^{(d)}:\{B_1^-,E_+^+\}$;\newline $A_2^{(d)}:\{E_-^-\}$;\newline $B_y^{(d)}:\{E_-^+\}$ \\
    \addlinespace[2pt]
    Off-axis, $\alpha=0$ & $C_2$ & $3A^{(d)}+3B^{(d)}$ & $A^{(d)}:\{B_1^+,E_+^-,E_-^-\}$;\newline $B^{(d)}:\{B_1^-,E_+^+,E_-^+\}$ \\
    \addlinespace[2pt]
    Diagonal, $\alpha=0$ & $C_{2v}$ & $A_1^{(d)}+2A_2^{(d)}$\newline $+B_{\parallel}^{(d)}+2B_{\perp}^{(d)}$ & $A_1^{(d)}:\{E_{\parallel}^-\}$;\newline $A_2^{(d)}:\{B_1^+,E_{\perp}^-\}$;\newline $B_{\parallel}^{(d)}:\{E_{\parallel}^+\}$;\newline $B_{\perp}^{(d)}:\{B_1^-,E_{\perp}^+\}$ \\
    Axial, $\alpha\ne0$ & $C_s$ & $4A'+2A''$ & $A':\{B_{1,A},E_{+,A},B_{1,B},E_{+,B}\}$;\newline $A'':\{E_{-,A},E_{-,B}\}$ \\
    \addlinespace[2pt]
    Off-axis, $\alpha\ne0$ & $C_1$ & $6A$ & All six local coordinates share one sector \\
    \addlinespace[2pt]
    Diagonal, $\alpha\ne0$ & $C_s$ & $2A'+4A''$ & $A':\{E_{\parallel,A},E_{\parallel,B}\}$;\newline $A'':\{B_{1,A},E_{\perp,A},B_{1,B},E_{\perp,B}\}$ \\
    \bottomrule
  \end{tabular}
\end{table}

Here $\xi$ labels an irrep distinct from $\zeta$. Repeated copies of the same irrep may mix. Table~\ref{tab:full-dimer-decomposition} resolves these multiplicity spaces. The site-exchange coordinates are $q^\pm=(q_A\pm q_B)/\sqrt2$, where $q_A$ and $q_B$ are the same local-mode amplitudes on the two sites; superscript $(d)$ distinguishes dimer irreps from local-family labels. Writing $\mathrm e$ for the identity operation and $\sigma_y$ for reflection about the $y$ axis, in the order $(\mathrm e,C_2,\sigma_x,\sigma_y)$, $B_x^{(d)}$ and $B_y^{(d)}$ have characters $(1,-1,1,-1)$ and $(1,-1,-1,1)$, respectively. The reflections $\sigma_\parallel$ and $\sigma_\perp$ are about the defect-connection diagonal and its perpendicular. The diagonal partners $E_\parallel,E_\perp$ are even and odd under reflection about the defect-connection diagonal; $B_\parallel^{(d)},B_\perp^{(d)}$ have the same respective characters in the order $(\mathrm e,C_2,\sigma_\parallel,\sigma_\perp)$.

The normalized residuals $\epsilon_C$ and $\epsilon_{\mathrm{forbid}}$ measure symmetry covariance and the size of symmetry-forbidden blocks, respectively, in the Frobenius norm $\|\cdot\|_{\mathrm F}$:
\begin{equation}
 \begin{aligned}
 \epsilon_C(g,\alpha)&=\frac{\|\bm D_\mu^{\mathrm H}(g)\bm C_\mu(\alpha)\bm D_\mu(g)-\bm C_\mu(\alpha)\|_{\mathrm F}}{\|\bm C_\mu(\alpha)\|_{\mathrm F}},\\
 \epsilon_{\mathrm{forbid}}(\alpha)&=\frac{\left[\sum_{\zeta\ne\xi}\|(\bm P_\zeta^\mu)^{\mathrm H}\bm C_\mu(\alpha)\bm P_\xi^\mu\|_{\mathrm F}^2\right]^{1/2}}{\|\bm C_\mu(\alpha)\|_{\mathrm F}}.
 \end{aligned}
 \label{eq:app-forbidden-block-residual}
\end{equation}

Over the PMMA--aluminum sweep, the maxima of $\epsilon_C$ ($\epsilon_{\mathrm{forbid}}$) are $3.64\times10^{-12}$ ($1.82\times10^{-12}$) axially and $1.01\times10^{-11}$ ($5.07\times10^{-12}$) diagonally. The nominal off-axis values are $3.32\times10^{-12}$ and $1.66\times10^{-12}$; detuning leaves $C_1$, with no nontrivial forbidden-block condition.

\section{Nominal Thresholds, Physical Mass Balance, and Family Crossings}
\label{app:threshold-analysis}

The nominal models use fixed recovery and consistently projected physical mass. This appendix gives the four-coordinate reversal reduction, the physical family-mass balance, the uncoupled-family crossing calculation, and the branch-resolved localization-root search.

\subsection{Nominal Threshold Reduction and Coupling Variation}
\label{app:nominal-threshold-derivation}
\begin{figure}[!b]
  \centering
  \begin{minipage}[t]{0.48\linewidth}
    \centering
    \includegraphics[trim=0 7.5pt 0 0,clip,width=\linewidth,height=147.5pt,keepaspectratio]{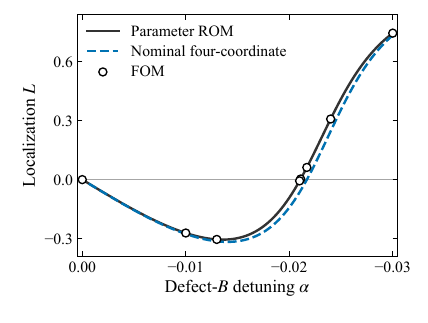}
    \par\smallskip{\small (a) Predicted localization and FOM comparisons.\par}
  \end{minipage}
  \hfill
  \begin{minipage}[t]{0.48\linewidth}
    \centering
    \includegraphics[trim=0 7.5pt 0 0,clip,width=\linewidth,height=147.5pt,keepaspectratio]{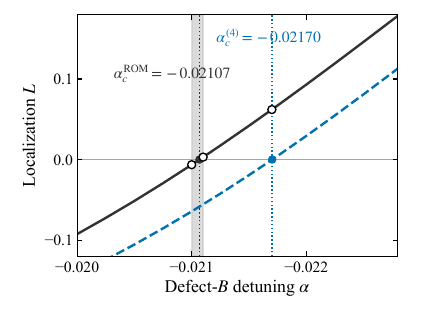}
    \par\smallskip{\small (b) Reversal thresholds.\par}
  \end{minipage}
\caption{Off-axis threshold prediction from the four-coordinate nominal model. FOM points are matched to the reference branch, with refined checks near the threshold in (b). The shaded interval brackets the FOM sign change.}
  \label{fig:nominal-threshold}
\end{figure}

The off-axis four-coordinate model retains the $B_1$ and $E_2$ channels at both sites. Its fixed physical basis selects columns 1, 2, 5, and 6 of the nominal recovery $\bm J(0)$ in the order of Eq.~\eqref{eq:fixed-channel-coordinates}. Figure~\ref{fig:nominal-threshold} shows its continued branch and the full-order checks at the threshold.

Write $\bm D=\bm K_4^0+\alpha\bm K_{4B}-\lambda\bm M_4$ in two-dimensional family blocks $b=B_1$ and $e=E_2$. For nonsingular $\bm D_{bb}$, eliminating the singlet coordinates gives
\begin{equation}
 \begin{aligned}
 \bm{\mathcal R}&=-\bm D_{bb}^{-1}\bm D_{be},&
 \bm{\mathcal S}&=\bm D_{ee}+\bm D_{eb}\bm{\mathcal R},\\
 \bm T_4&=\begin{bmatrix}\bm{\mathcal R}\\\bm I_2\end{bmatrix},&
 \bm N_{\mathrm{eff}}&=\bm T_4^{\mathrm H}\bm N_4\bm T_4.
 \end{aligned}
 \label{eq:nominal-threshold-schur}
\end{equation}

The nonzero reversal threshold satisfies the semi-analytical conditions
\begin{equation}
 \begin{gathered}
 \det\bm{\mathcal S}(\alpha_c,\lambda_c)=0,\qquad
 \bm e^{\mathrm H}\bm N_{\mathrm{eff}}(\alpha_c,\lambda_c)\bm e=0,\\
 \bm e\in\ker\bm{\mathcal S}\setminus\{\bm0\}.
 \end{gathered}
 \label{eq:nominal-reversal-threshold}
\end{equation}

Here $\bm e\in\mathbb C^2$ contains the retained $E_2$ coordinates, $\bm I_2$ is the $2\times2$ identity matrix, and $\ker\bm{\mathcal S}$ denotes the null space of the Schur complement. The map $\bm T_4$ reconstructs the complete four-coordinate vector from $\bm e$, and $\bm N_{\mathrm{eff}}$ is its effective defect-mass difference form. For the real symmetric matrices used here, $\bm e=(\mathcal S_{12},-\mathcal S_{11})^{\mathrm T}$ reduces Eq.~\eqref{eq:nominal-reversal-threshold} to two rational equations in $(\alpha_c,\lambda_c)$ wherever this vector is nonzero. The root is selected on the continued target branch, excluding nominal site balance at $\alpha=0$.

The coupling scaling in Eq.~\eqref{eq:coupling-strength-intervention} samples $\gamma=0.01$, 0.02, 0.05, and 0.1--1 in increments of 0.1. The off-axis root moves from $-0.02170265$ at $\gamma=1$ to $-0.023261$ at $0.1$ and $-0.023400$ at $0.01$, approaching the uncoupled crossing at $-0.02341527$. The corresponding root-slope magnitudes are 96.7, 1097, and 11091. At the original four-coordinate root, $\bm D_{bb}$ is nonsingular and the nearest frequency separation is \SI{1.559}{Hz}.

Over the weak-coupling samples the root-slope magnitude varies approximately as $1/\gamma$. The coupling-strength study in Fig.~\ref{fig:coupling-strength} resolves the narrow avoided crossings of the prescribed four-coordinate pencil by refined continuation. At $\gamma=0.01$, the minimum frequency gap is \SI{0.01519}{Hz}; 12000 detuning intervals give a minimum successive physical MAC of 0.999677, and doubling this resolution changes $\alpha_c$ by less than $10^{-12}$. At $\gamma=0$, invariant-family identity resolves the exact crossing.

\subsection{Projection onto the Crossing Pair and Corrections to the Threshold}
\label{app:pair-projection}

Write $\bm V=[\bm u_b,\bm u_e]$ and $\bm M_2=\bm V^{\mathrm{T}}\bm M\bm V$ in the four-dimensional mirror sector. For a complete mass-normalized eigenvector $\bm y$, the projection and retained mass fraction are
\begin{equation}
 \bm y_\parallel=\bm V\bm M_2^{-1}\bm V^{\mathrm{T}}\bm M\bm y,\qquad
 \Pi_2=\bm y_\parallel^{\mathrm{T}}\bm M\bm y_\parallel,\qquad
 \bm y_\perp=\bm y-\bm y_\parallel.
\end{equation}

The projected state has $L=+0.1003$ in PMMA--aluminum and $-0.4566$ in aluminum--tungsten at their crossings. Mass orthogonality gives $\bm y_\parallel^{\mathrm{T}}\bm M\bm y_\perp=0$, while the observable contains $2\bm y_\parallel^{\mathrm{T}}\bm N\bm y_\perp$. In PMMA--aluminum the three contributions from $\bm y_\parallel$, the interference term, and $\bm y_\perp$ are $+0.07169$, $+0.05721$, and $-0.00202$. In both cases the paired two-state target is the upper projected mode at the crossing, selected by the squared mass overlap with the complete target state.

For comparison with the frozen crossing pair we restore either omitted mode or correct the pair once using both omitted modes. In the basis of the four uncoupled modes at $s_\times$, write $\bm A=\bm K(s_\times)-\lambda_\times\bm M$ in blocks $P$ (crossing pair) and $R$ (remaining pair). With nonsingular $\bm A_{RR}$ the corrected basis has coefficients $[\bm I;\,-\bm A_{RR}^{-1}\bm A_{RP}]$. This basis is frozen before projecting the stiffness, mass, and both cell-mass operators over the softening interval. A separate local estimate uses the full nominal sector at the crossing:
\begin{equation}
 s_{\mathrm{lin}}=s_\times-\frac{D(s_\times)}{D'(s_\times)},\qquad
 D'=2\,\bm y'^{\mathrm{T}}\bm N\bm y.
\end{equation}

For a simple eigenvalue with fixed $\bm M$, the derivative follows by expanding $\bm y'$ over the other mass-normalized sector eigenvectors $\bm y_j$, with coefficients $\bm y_j^{\mathrm{T}}(-\bm K_B)\bm y/(\lambda-\lambda_j)$. The derivatives with respect to dimensionless $s$ at the two crossings are $+61.42$ and $+21.52$. Table~\ref{tab:pair-root-comparison} compares the resulting softening levels.

\begin{table}[!b]
\centering
\caption{Nonzero reversal thresholds on the two diagonal target branches. Each entry gives the percentage softening of defect $B$.}
\label{tab:pair-root-comparison}
\begin{tabular}{lrr}
\toprule
Model or estimate & PMMA--aluminum & Aluminum--tungsten\\
\midrule
Frozen crossing pair & 2.0287 & 2.8856\\
Pair plus remaining singlet mode & 2.0290 & 2.4523\\
Pair plus remaining doublet-partner mode & 1.9482 & 2.5632\\
Corrected frozen pair & 2.0047 & 2.3227\\
Local linear estimate, complete sector & 1.9784 & 2.2616\\
Direct root, complete sector & 1.9648 & 2.2425\\
Independent full-order bracket & 1.950--1.960 & 2.235--2.240\\
\bottomrule
\end{tabular}
\end{table}

\subsection{Physical Family Contributions at the Reversal Threshold}
\label{app:family-mass-balance}

For the complete nominal six-coordinate model, set $\bm V_6^{\mathrm{phy}}=\bm J(0)$ and define $\bm M_6$, $\bm Q_{A6}$, and $\bm Q_{B6}$ by the projections used in Eq.~\eqref{eq:nominal-projected-operators}. Let $\bm N_6=\bm Q_{A6}-\bm Q_{B6}$ and $\bm Q_6=\bm Q_{A6}+\bm Q_{B6}$. Partition the mass-normalized state into $\bm y_b\in\mathbb C^2$ for $B_1$ and $\bm y_e\in\mathbb C^4$ for the complete $E$ family, with $\bm y^{\mathrm H}\bm M_6\bm y=1$. Define
\begin{equation}
\begin{aligned}
 n_b&=\bm y_b^{\mathrm H}\bm N_{6,bb}\bm y_b,&
 n_e&=\bm y_e^{\mathrm H}\bm N_{6,ee}\bm y_e,\\
 n_x&=2\operatorname{Re}(\bm y_b^{\mathrm H}\bm N_{6,be}\bm y_e),&
 q_x&=2\operatorname{Re}(\bm y_b^{\mathrm H}\bm Q_{6,be}\bm y_e),\\
 q_b&=\bm y_b^{\mathrm H}\bm Q_{6,bb}\bm y_b,&
 q_e&=\bm y_e^{\mathrm H}\bm Q_{6,ee}\bm y_e.
\end{aligned}
\label{eq:family-mass-contributions}
\end{equation}

The subscripts $b,e$ label the $B_1$ and $E$ families. Their signed contributions are evaluated on the continued modal state. Retaining both cross terms gives
\begin{equation}
 L_6=\frac{n_b+n_e+n_x}{q_b+q_e+q_x}.
 \label{eq:family-mass-quotient}
\end{equation}

For $q_b,q_e>0$, define the family-block defect-mass share $p=q_b/(q_b+q_e)$, the spatial biases $\ell_b=n_b/q_b$, $\ell_e=n_e/q_e$, and $\xi=n_x/(q_b+q_e)$. When $\ell_b\ne\ell_e$,
\begin{equation}
 \begin{aligned}
  p_* &= \frac{-\ell_e-\xi}{\ell_b-\ell_e},\\
  L_6&=\frac{q_b+q_e}{q_b+q_e+q_x}(\ell_b-\ell_e)(p-p_*).
 \end{aligned}
 \label{eq:family-mass-balance-share}
\end{equation}

This identity expresses the physical numerator balance in terms of unequal family biases. Both $p$ and $p_*$ vary along the branch; the eigenproblem determines whether their trajectories meet. The share $p$ is normalized by the sum of the two family-block defect masses. The cross-family mass $q_x$ is retained separately.

\begin{table}[!b]
 \centering
 \caption{PMMA--aluminum physical contributions at the complete nominal roots in Table~\ref{tab:threshold-accuracy}. All entries use unit total projected modal mass.}
 \label{tab:family-mass-balance}
 \begin{tabular}{l*{2}{S[group-digits=none,table-format=+1.6,retain-explicit-plus]}}
  \toprule
  Quantity & {Off-axis} & {Diagonal} \\
  \midrule
  $n_b$ & +0.252843 & +0.266696\\
  $n_e$ & -0.253738 & -0.267202\\
  $n_x$ & +0.000895 & +0.000506\\
  $q_x$ & -0.003330 & -0.003996\\
  $p=p_*$ & 0.372768 & 0.372280\\
  $\eta_A=\eta_B$ & 0.356565 & 0.357126\\
  \bottomrule
 \end{tabular}
\end{table}

Table~\ref{tab:family-mass-balance} shows that reversal is dominated by cancellation of opposite family contributions. Equal family-block defect masses, $p=1/2$, occur later along softening, at $\alpha=-0.0231373$ off axis and $-0.0220141$ diagonally. These locations differ from both the localization roots and the uncoupled family crossings. The four-coordinate off-axis model has its own root contributions, approximately $+0.262$ from $B_1$, $-0.263$ from $E_2$, and less than 0.001 from the cross term, corresponding to its retained $B_1$ and $E_2$ subspace.

Keeping the six-coordinate eigenvectors fixed while omitting only $n_x$ moves the roots by $-1.36\times10^{-5}$ and $-9.71\times10^{-6}$, respectively. The correction to the localization index is small, whereas removing cross-family stiffness and mass blocks changes the mode shapes themselves, as shown by the controls in Sec.~\ref{sec:fixed-branch-results}.

\begin{figure}[!b]
  \centering
  \begin{minipage}[t]{0.48\linewidth}
    \centering
    \includegraphics[trim=0 7.5pt 0 0,clip,width=\linewidth,height=147.5pt,keepaspectratio]{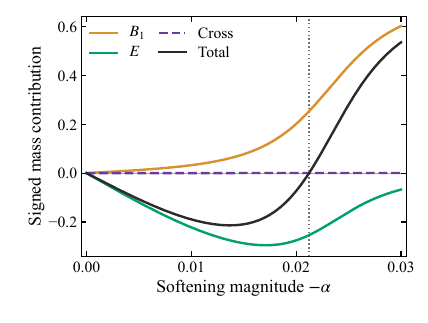}
    \par\smallskip{\small (a) Off-axis mass contributions.\par}
  \end{minipage}
  \hfill
  \begin{minipage}[t]{0.48\linewidth}
    \centering
    \includegraphics[trim=0 7.5pt 0 0,clip,width=\linewidth,height=147.5pt,keepaspectratio]{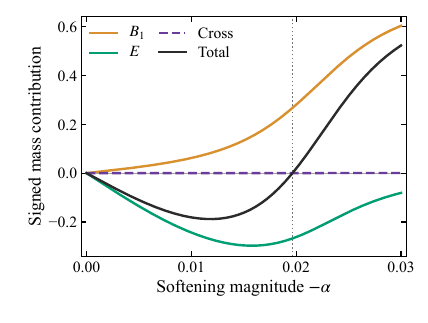}
    \par\smallskip{\small (b) Diagonal mass contributions.\par}
  \end{minipage}
  \par\smallskip
  \begin{minipage}[t]{0.48\linewidth}
    \centering
    \includegraphics[trim=0 7.5pt 0 0,clip,width=\linewidth,height=147.5pt,keepaspectratio]{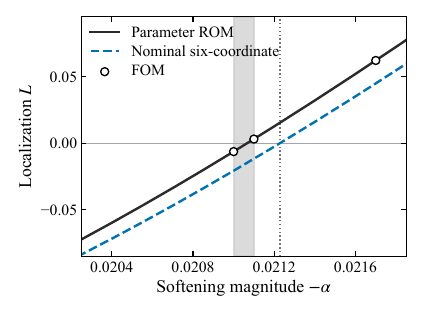}
    \par\smallskip{\small (c) Off-axis threshold.\par}
  \end{minipage}
  \hfill
  \begin{minipage}[t]{0.48\linewidth}
    \centering
    \includegraphics[trim=0 7.5pt 0 0,clip,width=\linewidth,height=147.5pt,keepaspectratio]{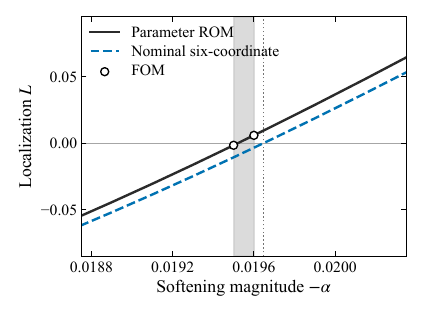}
    \par\smallskip{\small (d) Diagonal threshold.\par}
  \end{minipage}
  \caption{Physical mass balance on the selected softening branches. (a,b) Signed defect-mass contributions of the nominal six-coordinate model. (c,d) Predicted localization near the threshold, compared with the reference ROM and independent FOM checks. Vertical dotted lines mark nominal predictions; shaded bands bracket FOM sign changes.}
  \label{fig:family-mass-balance}
\end{figure}

Figure~\ref{fig:family-mass-balance} traces the PMMA--aluminum family contributions through reversal. At the off-axis reversal, the $B_1$ and $E$ biases are $+0.947$ and $-0.565$, respectively, giving balance at $p=37.28\%$. For comparison with overall family character, $\Pi_g$ denotes the full-space physical-mass projection in Eq.~\eqref{eq:family-projection-score} of the supplementary material onto the nominal physical basis $\bm U_g$ of family $g$; because the translated family spaces are not mutually mass-orthogonal, these projections need not sum to one, unlike the family-block defect-mass share $p$ in Eq.~\eqref{eq:family-mass-balance-share}. Along the selected nominal softening branches, $\Pi_{B_1}$ increases from 0.0362 to 0.8289 off axis and from 0.0415 to 0.8187 diagonally, while at the reversal roots $\Pi_E$ remains 0.6388 and 0.6386. Over 600 continuation intervals on $[-0.03,0]$, the minimum target-branch physical MAC between the nominal six-coordinate model and the parameter ROM is 0.99960 off axis and 0.99975 diagonally (see Sec.~\ref{app:primary-rom-validation} of the supplementary material for the corresponding FOM threshold checks).

\subsection{Uncoupled Family Crossings}
\label{app:family-crossings}

We locate crossings by suppressing the cross-family dynamical blocks in the fixed nominal basis. For the PMMA--aluminum dimer we use the two-site $B_1$ block and the four-coordinate $E$ block. In the diagonal aluminum--tungsten dimer, the interacting states belong to the $B_2$ and mirror-even $E$ site pairs, each of dimension two. For each block $j$ of dimension $d_j$, let $\bm K_j^0$, $\bm K_j^B$, and $\bm M_j$ denote its nominal stiffness, detuning stiffness, and mass. Whitening the mass within this block gives
\begin{equation}
 \begin{aligned}
  \bm H_j(\alpha)&=\bm H_j^0+\alpha\bm C_j,\\
  \bm H_j^0&=\bm M_j^{-1/2}\bm K_j^0\bm M_j^{-1/2},\\
  \bm C_j&=\bm M_j^{-1/2}\bm K_j^B\bm M_j^{-1/2}.
 \end{aligned}
 \label{eq:family-block-whitening}
\end{equation}

The uncoupled family spectra intersect at detunings $\alpha_\times$ given by the finite real generalized eigenvalues of
\begin{equation}
 [\bm H_i^0\otimes\bm I_{d_j}-\bm I_{d_i}\otimes\bm H_j^0]\bm z
 =-\alpha_\times[\bm C_i\otimes\bm I_{d_j}-\bm I_{d_i}\otimes\bm C_j]\bm z,
 \label{eq:uncoupled-family-crossing}
\end{equation}
for $\bm z\ne\bm0$ in the $d_id_j$-dimensional product space, where $\bm I_{d_i}$ and $\bm I_{d_j}$ are identity matrices of sizes $d_i$ and $d_j$. For normalized crossing states $\bm u_i$ and $\bm u_j$ with common eigenvalue $\lambda=(2\pi f)^2$ and cross-family stiffness and mass blocks $\bm K_{ij}(\alpha_\times)$ and $\bm M_{ij}$, the restored cross-family coupling strength is evaluated on the linearized frequency scale of Eq.~\eqref{eq:first-order-equivalent-frequency} of the supplementary material as $|\bm u_i^{\mathrm H}[\bm K_{ij}(\alpha_\times)-\lambda\bm M_{ij}]\bm u_j|/(8\pi^2 f)$. This equivalent coupling is \SI{5.53}{Hz} for the aluminum--tungsten diagonal pair. Table~\ref{tab:primary-layout-roots} gives the couplings and sign-change counts across the PMMA--aluminum layouts.

\begin{table}[!t]
  \centering
  \caption{Uncoupled family crossings and coupled localization reversals in the PMMA--aluminum system over $-0.03\leq\alpha\leq0.03$. The last column gives the number of nonzero sign changes and the number meeting the two-sided amplitude criterion $|L|\geq0.1$ of Sec.~\ref{sec:mechanism-ablation}. The reference-ROM branch counts agree with the nominal-model counts in every layout.}
  \label{tab:primary-layout-roots}
  \setlength{\tabcolsep}{4pt}
  \begin{tabular}{@{}llcc@{}}
    \toprule
    Layout & Uncoupled crossings $\alpha_\times$ & Coupling (Hz) & Zeros / substantial reversals \\
    \midrule
    Axial & None in the scan & --- & 0 / 0 \\
    Off-axis & $-0.01303,+0.01320$ & $\sim\!0.41$ & 8 / 6 \\
             & $-0.02342,+0.02393$ & 0.75--0.77 & \\
    Diagonal & $-0.02185,+0.02191$ & $\sim\!1$ & 4 / 4 \\
             & $-0.02279,+0.02291$ & $<10^{-9}$ & \\
    \bottomrule
  \end{tabular}
\end{table}

\subsection{Direct Localization Roots of Multimode Nominal Models}
\label{app:direct-localization-roots}
For each PMMA--aluminum layout, consistent Galerkin projection onto the fixed six-coordinate nominal basis $\bm V_d^{\mathrm{phy}}$ ($d=6$) defines $\bm K_d^0$, $\bm K_{dB}$, $\bm M_d$, $\bm N_d$, and $\bm Q_d$ as in Eqs.~\eqref{eq:nominal-projected-pencil} and~\eqref{eq:nominal-projected-operators}. Starting from the nominal eigenmodes on each half of the detuning interval, mass-weighted continuation brackets candidate zeros where $(\bm K_d^0+\alpha_c\bm K_{dB}-\lambda_c\bm M_d)\bm y=\bm0$ and $\bm y^{\mathrm H}\bm N_d\bm y=0$ with $\bm y^{\mathrm H}\bm Q_d\bm y>0$ (excluding $\alpha=0$). Refining the grid from 600 to 1200 intervals per half-interval leaves all root counts unchanged and shifts root locations by less than $2\times10^{-14}$, agreeing with direct root solves to within $9\times10^{-15}$; no stationary zero contacts occur. Evaluating the adjacent sign lobes excludes two off-axis zeros whose smaller lobes remain below $|L|=0.1$.

In the diagonal PMMA--aluminum layout, the symmetry-isolated $(B_1,E_1)$ four-coordinate block and $E_2$ two-coordinate block have relative off-block norms below $4.7\times10^{-12}$, allowing block-resolved continuation through near-coincident frequencies. The six-coordinate off-axis and diagonal thresholds in Table~\ref{tab:threshold-accuracy} lie on the branches continuing the fourth and third nominal modes in ascending frequency order.

\bibliographystyle{elsarticle-num}
\bibliography{references}

\clearpage
\section*{Supplementary Material}
\addcontentsline{toc}{section}{Supplementary Material}
\counterwithout{equation}{section}
\setcounter{section}{0}
\setcounter{equation}{0}
\setcounter{table}{0}
\setcounter{figure}{0}
\renewcommand{\thesection}{S\arabic{section}}
\renewcommand{\theequation}{S\arabic{equation}}
\renewcommand{\thetable}{S\arabic{table}}
\renewcommand{\thefigure}{S\arabic{figure}}
\renewcommand{\theHsection}{supp.\arabic{section}}
\renewcommand{\theHequation}{supp.\arabic{equation}}
\renewcommand{\theHtable}{supp.\arabic{table}}
\renewcommand{\theHfigure}{supp.\arabic{figure}}
\captionsetup[subfigure]{font=small,labelformat=parens,labelsep=space,justification=centering,singlelinecheck=false}

Equations, tables, and figures with an S prefix belong to the supplementary material below. Other numbers refer to the main article.

\section{Numerical Settings, Retained Modal Sets, and Reference Validation}
\label{app:numerical-validation}

For the PMMA--aluminum system, projection onto the retained modal set checks the translated basis, and independent full-order model (FOM) comparisons assess the parameter-enriched reference. For the aluminum--tungsten dimer, the target branch is continued directly in the FOM.

\subsection{Discretization and Reduced-Space Parameters}
\label{app:rom-settings}
Both systems use $60\times60$ plane-strain four-node quadrilaterals per cell, $2\times2$ Gauss integration, and consistent mass. Each binary-mask pixel specifies the material of one element. The periodic boundary conditions and physical detuning are defined in main-text Sec.~\ref{sec:numerical-setup}. For the PMMA--aluminum $5\times5$ supercell at $\Gamma$, the shift-and-invert FOM eigensolver computes 20 eigenpairs, with at most 1000 iterations and an eigenvalue tolerance of $10^{-10}$. Reduced Hermitian generalized problems use dense direct eigensolution. For the $q$th enrichment system, $(\bm K_*-\lambda_{\mathrm{enr}}\bm M)\bm x_q=\bm b_q$, $\bm x_q$ is the solution and $\bm b_q$ the right-hand side. The normalized eigenpair and enrichment-solve residuals are
\begin{equation}
 \begin{aligned}
 r_j^{\mathrm{eig}}&=\frac{\|\bm K\bm q_j-\lambda_j\bm M\bm q_j\|_2}{\|\bm K\bm q_j\|_2+|\lambda_j|\|\bm M\bm q_j\|_2+\varepsilon},\\
 r_q^{\mathrm{solve}}&=\frac{\|\bm K_*\bm x_q-\lambda_{\mathrm{enr}}\bm M\bm x_q-\bm b_q\|_2}{\|\bm K_*\bm x_q\|_2+|\lambda_{\mathrm{enr}}|\|\bm M\bm x_q\|_2+\|\bm b_q\|_2+\varepsilon},
 \end{aligned}
 \label{eq:app-eigen-residual}
\end{equation}
where $\varepsilon>0$ prevents division by zero, $\lambda_{\mathrm{enr}}=(2\pi f_{\mathrm{enr}})^2$ is the enrichment shift at frequency $f_{\mathrm{enr}}$, and $\bm K_*=\bm K_0+\alpha_*\bm K_B$ is the stiffness at the enrichment anchor $\alpha_*$. Shifted-solve residuals are at most $10^{-8}$. For every mass-orthonormal basis $\bm V$, Galerkin projection gives $\widehat{\bm K}(\alpha)=\bm V^{\mathrm H}(\bm K_0+\alpha\bm K_B)\bm V$ and $\widehat{\bm M}=\bm V^{\mathrm H}\bm M\bm V$.

Table~\ref{tab:case-specific-rom-parameters} lists the PMMA--aluminum reduced-space settings, including the enrichment and condensation frequencies $f_{\mathrm{enr}}$ and $f_{\mathrm{cond}}$. The frequency-enriched space uses the six translated local modes and three mass-response blocks from Eq.~\eqref{eq:host-response-enrichment}. The parameter-enriched space uses six nominal coupled Ritz vectors and three stiffness-response blocks. Section~\ref{app:secondary-rom-validation} gives the separate fixed-shift construction for the aluminum--tungsten dimer.

\begin{table}[!b]
  \centering
  \caption{Reduced-space construction parameters for the PMMA--aluminum system. The rank tolerance is relative to the largest mass-Gram eigenvalue.}
  \label{tab:case-specific-rom-parameters}
  \setlength{\tabcolsep}{3pt}
  \begin{tabular}{L{0.42\linewidth}L{0.48\linewidth}}
    \toprule
    Parameter & PMMA--aluminum ($B_1+E$) system \\
    \midrule
    Fixed independent-channel order & $B_1,E_1,E_2$ \\
    Enrichment and condensation shifts & $f_{\mathrm{enr}}=f_{\mathrm{cond}}=\SI{1200}{Hz}$ \\
    Reference seeds & Six nominal coupled Ritz modes from the frequency-enriched model \\
    Reference response blocks & Three six-column blocks \\
    Reference ROM dimension & 24 \\
    Frequency-enriched dimension & 24 \\
    Relative mass-Gram rank tolerance & $10^{-10}$ \\
    \bottomrule
  \end{tabular}
\end{table}

\subsection{Retained Modal Sets}
\label{app:family-retention}
Let $\bm F^{\mathrm{phy}}$ collect linearly independent full-space displacement vectors of the retained modal set as its columns. For the subspace $\mathcal F=\operatorname{range}(\bm F^{\mathrm{phy}})$, the normalized physical-mass projection is
\begin{equation}
  \Pi_{\mathcal{F},r}
  =\frac{\bm{u}_r^{\mathrm{H}}\bm{M}^{\mathrm{phy}}\bm{F}^{\mathrm{phy}}
  \left[(\bm{F}^{\mathrm{phy}})^{\mathrm{H}}\bm{M}^{\mathrm{phy}}\bm{F}^{\mathrm{phy}}\right]^{-1}
  (\bm{F}^{\mathrm{phy}})^{\mathrm{H}}\bm{M}^{\mathrm{phy}}\bm{u}_r}
  {\bm{u}_r^{\mathrm{H}}\bm{M}^{\mathrm{phy}}\bm{u}_r}.
  \label{eq:family-projection-score}
\end{equation}

The PMMA--aluminum single-defect modal set comprises a nondegenerate $B_1$ mode (singlet) at \SI{1198.751}{Hz} and a degenerate $E$ mode pair (doublet) at \SI{1205.721}{Hz} within the \SIrange{900.326}{1381.164}{Hz} host bandgap. All three modes persist along the complete dispersion path, remain at least \SI{174.938}{Hz} from the gap edges, and have defect-cell modal mass fraction at least 0.702. This confirms appreciable confinement together with a finite field outside the defect cell; cross-defect overlap is retained explicitly in the Gram matrix.

Membership in the single-defect modal set along $\Gamma$--$\mathrm X$--$\mathrm M$--$\Gamma$ uses this projection onto the $\Gamma$-point modal set and the defect-cell mass fraction. For the PMMA--aluminum system over the full dimer detuning sweep, the minimum combined defect-cell mass fractions $\eta_A+\eta_B$ are 0.689, 0.700, and 0.698 in the axial, off-axis, and diagonal layouts, respectively. The corresponding minimum projection scores onto the translated modal set are 0.9961, 0.9983, and 0.9999.

\subsection{PMMA--Aluminum Reference Validation}
\label{app:primary-rom-validation}
The original PMMA--aluminum FOM set contains $\alpha=-0.03$, $-0.01$, $0$, $0.01$, $0.03$ in all layouts, plus additional verification points $-0.024,-0.013,0.013,0.023$ off axis and $\pm0.022$ diagonally: 21 states and 126 retained-mode comparisons. Joint cluster exclusions leave 112 common individual modes. Table~\ref{tab:detuning-validation} compares the signed-localization errors of the parameter- and frequency-enriched ROMs on this common subset. The parameter-ROM frequency, modal-mass-fraction, and subspace-angle statistics use the complete six-mode sets. The parameter-ROM errors are small relative to the independent-channel discrepancies and are lower in signed localization than those of the frequency-enriched space in all layouts.

\begin{table}[!b]
  \centering
  \caption{ROM--FOM validation for the PMMA--aluminum system. Signed-localization errors use 112 common unclustered modes; other metrics use the complete six-mode sets. Both enriched spaces use the same eligible modes for $L$.}
  \label{tab:detuning-validation}
  \setlength{\tabcolsep}{5pt}
  \begin{tabular}{@{}L{0.43\linewidth}rrr@{}}
    \toprule
    Metric & \multicolumn{1}{c}{Axial $(2,0)$} & \multicolumn{1}{c}{Off-axis $(2,1)$} & \multicolumn{1}{c}{Diagonal $(2,2)$} \\
    \midrule
    FOM comparison cases & 5 & 9 & 7 \\
    Maximum relative frequency error & $1.831\times10^{-7}$ & $1.367\times10^{-7}$ & $1.333\times10^{-7}$ \\
    \begin{tabular}[t]{@{}l@{}}Maximum defect-cell\\modal-mass-fraction error\end{tabular} & $1.333\times10^{-4}$ & $1.360\times10^{-4}$ & $1.408\times10^{-4}$ \\
    Maximum six-mode principal angle & 0.0591$^\circ$ & 0.0540$^\circ$ & 0.0555$^\circ$ \\
    Eligible individual modes for $L$ & 30 & 50 & 32 \\
    Maximum $|L_{\mathrm{ROM}}-L_{\mathrm{FOM}}|$ & $2.134\times10^{-4}$ & $1.137\times10^{-4}$ & $1.017\times10^{-4}$ \\
    \begin{tabular}[t]{@{}l@{}}Frequency-enriched 24D:\\maximum $|L-L_{\mathrm{FOM}}|$\end{tabular} & 0.007485 & 0.030337 & 0.017569 \\
    \bottomrule
  \end{tabular}
\end{table}

Three refined off-axis FOM checks near the reversal threshold give $L_{\mathrm{FOM}}=+0.0620885$ at $\alpha=-0.02170$, $+0.0030396$ at $-0.02110$, and $-0.0063733$ at $-0.02100$. The last two values bracket the reversal reported as $s=-\alpha$ in main-text Table~\ref{tab:threshold-accuracy}, with linear interpolation estimating $\alpha_c^{\mathrm{FOM}}\approx-0.02106771$. The reference-ROM root is $-0.02106786$, and the maximum four-coordinate localization discrepancy over 121 samples in $[-0.03,0]$ is 0.0654. Across all 18 retained modes at these three states, the maximum FOM eigen-residual is $2.1\times10^{-11}$ and the minimum ROM--FOM MAC is $0.9999999511$; on the target branch, the nearest FOM modal separation exceeds \SI{1.47}{Hz} and the MAC is at least $0.999999999995$.

Two refined diagonal FOM checks give $L_{\mathrm{FOM}}=+0.00583773$ at $\alpha=-0.01960$ and $-0.00162100$ at $-0.01950$, defining the diagonal PMMA--aluminum bracket reported as $s=-\alpha$ in main-text Table~\ref{tab:threshold-accuracy} (linearly interpolated root: $-0.01952173$). At 2.200\% softening, the nearest archived full-order sample to the 2.18502\% nominal crossing on this branch, $\eta_A=0.434898$ and $\eta_B=0.287569$, giving $L=+0.203925$. The matched mode has frequency \SI{1199.8763}{Hz} and mass-weighted overlap above 0.9999999999. Across the 12 retained modes at these two states, the maximum FOM eigen-residual is $5.3\times10^{-11}$ and the minimum ROM--FOM MAC is $0.999999970229$. On the target branch (continuing the third nominal mode in ascending frequency order), the nearest FOM mode separation within the retained set exceeds \SI{0.56}{Hz}, the relative frequency gap remains above $4.67\times10^{-4}$ (well above $g_{\mathrm{cl}}=10^{-4}$), and the ROM--FOM MAC is at least $0.999999999968$.

To test the effect of supercell size, we compared the off-axis branch associated with the lower-frequency nominal $E_2$ site-pair mode in the $5\times5$ and $7\times7$ supercells while keeping the defect geometry, offset $(2,1)$, and mesh per cell fixed. The branch reverses from the softened defect $B$ toward $A$ in both supercells, with full-order solutions bracketing the reversal at 2.10\%--2.11\% softening in the $5\times5$ supercell and at 2.4\%--2.6\% in the $7\times7$ supercell. For each of the two bracketing $7\times7$ full-order calculations at $\Gamma$, we computed 12 eigenpairs using shift-and-invert with a shift frequency of \SI{1200}{Hz} and an eigensolver tolerance of $10^{-11}$; across these two points, $L$ changes from $-0.03636$ to $+0.16115$, and the two full-order mode shapes have a mass-weighted MAC of 0.9842.

\subsection{Aluminum--Tungsten Defect and Full-Order Validation}
\label{app:secondary-rom-validation}
The binary mask assigns tungsten to elements whose centers lie inside the square or disks specified in the main article. The resulting tungsten area fraction is $1024/3600$. In the $5\times5$ periodic supercell, the single defect occupies $(2,2)$, and the diagonal dimer adds an identical defect at $(4,4)$.

The single-defect dispersion uses four equal intervals on each segment of $\Gamma$--$\mathrm X$--$\mathrm M$--$\Gamma$, giving 13 path points with the $\Gamma$ solution reused at the end. The Bloch phase vector $\bm\theta=5a\bm k$ at the path vertices is $(0,0)$, $(\pi,0)$, $(\pi,\pi)$, and $(0,0)$. Each eigensolution uses 12 modes per parity of the mirror that preserves the wavevector modulo the reciprocal lattice. The shift is the mean of the $\Gamma$-point $B_2$ and $E$ frequencies. We match successive nodal periodic envelopes $e^{-i\bm k\cdot(\bm r-\bm r_c)}\bm u_{\bm k}(\bm r)$ in the fixed $\Gamma$ mass metric, where $\bm r$ is the nodal position vector, $\bm r_c$ is the defect center, and $\bm u_{\bm k}(\bm r)$ is the Bloch displacement field. Modes within \SI{0.001}{Hz} are grouped by adjacent frequency gaps and compared as subspaces using mass-weighted principal angles. The sampled frequencies span \SIrange{2301.7762}{2302.3477}{Hz} for the branch issuing from $B_2$ and \SIrange{2319.4764}{2319.8780}{Hz} for the two branches issuing from $E$. The minimum separation from another computed mode is \SI{205.413}{Hz}, and the largest full-space eigenpair residual among the tracked modes is $2.30\times10^{-13}$.

For the dimer, $\bm W$ contains both translations of the $\Gamma$-point $B_2$ singlet at \SI{2301.776246}{Hz} and of both $E$ partners at \SI{2319.792348}{Hz}. The partners $E_{\parallel}$ and $E_{\perp}$ are even and odd under reflection about the inter-defect diagonal, respectively. The nominal shift $f_\sigma=\SI{2310.7843}{Hz}$ is the mean of the singlet and doublet frequencies. Host recovery eliminates the full finite-element complement of $\bm W$ orthogonal in the $\bm M$ metric. Let $\bm A_\sigma=\bm K_0-(2\pi f_\sigma)^2\bm M$ and $\bm S=\bm W^{\mathrm H}\bm M\bm W$. Six shifted solves give the recovered basis $\bm U$ through
\begin{equation}
 \bm A_\sigma\bm X=\bm M\bm W,
 \qquad
 \bm U=\bm X(\bm W^{\mathrm H}\bm M\bm X)^{-1}\bm S.
 \label{eq:alw-full-complement-recovery}
\end{equation}

This construction satisfies $\bm W^{\mathrm H}\bm M\bm U=\bm S$ and $\bm Z^{\mathrm H}\bm A_\sigma\bm U=\bm 0$ for every $\bm Z$ with $\bm W^{\mathrm H}\bm M\bm Z=\bm 0$. The relative residuals of the shifted solve and coordinate constraint are $4.97\times10^{-11}$ and $2.34\times10^{-15}$, using the Frobenius norms of $\bm M\bm W$ and $\bm S$ for normalization. The matrices $\bm K_0$, $\bm K_B$, $\bm M$, $\bm M_A$, and $\bm M_B$ are all projected by congruence with the same fixed $\bm U$. Here $\bm M_A$ and $\bm M_B$ are the element-assembled defect-cell masses in Bloch-reduced coordinates. The four even coordinates form the $(B_2,E_{\parallel})$ block, and the two odd coordinates form the $E_{\perp}$ block. To locate the uncoupled spectral intersection, we set the $B_2$--$E_{\parallel}$ stiffness and mass blocks to zero and intersect the two site-pair spectra in the even sector.

The independent FOM calculation uses the exact $\Gamma$ mirror-even subspace of the finite-element pencil $\bm K_0+\alpha\bm K_B-\lambda\bm M$. Ten eigenpairs near \SI{2319.673732}{Hz} are computed using sparse shift-and-invert with tolerance $10^{-9}$. The 36 detunings comprise $\alpha=-j/1000$ for $j=0,\ldots,30$, together with $-0.02235$, $-0.02240$, $-0.02245$, $-0.02250$, and the nominal prediction ($\alpha_c\approx-0.022425$). The fourth nominal mode of the retained set is the third even mode within that set. Its physical mass MAC with the corresponding nominal FOM mode is 0.9999999984. At each softening step, the neighboring FOM mode with the highest physical mass MAC continues this branch. The smallest successive MAC is 0.998118, the smallest gap to another even mode is \SI{6.051}{Hz}, and the largest full-space relative residual on the branch is $6.65\times10^{-13}$.

Linear interpolation between the FOM solutions at $\alpha=-0.02240$ and $-0.02235$ gives $\alpha_c^{\mathrm{FOM}}\simeq-0.02236584$. The sign-lobe test excludes the nominal balance at $\alpha=0$ and requires both adjacent lobes to reach $|L|=0.1$. Across all 36 points the nominal model differs from the tracked FOM branch by at most 0.003925 in $L$ and \SI{0.040763}{Hz} in frequency.

For the dynamical control, the off-diagonal $B_2$--$E_{\parallel}$ stiffness and mass blocks are multiplied by a common factor $\gamma$. At $\alpha=0$, the state is matched in the original projected mass metric over 200 equal steps from $\gamma=1$ to 0. The matched state is then continued within $E_{\parallel}$ as defect $B$ softens. The full projected cell-mass forms give $L=-0.97864$ at $\alpha=-0.03$ in the decoupled limit. For the control at fixed mode shapes, the two cell masses are evaluated using only their within-family blocks. The zero of the resulting index lies at $\alpha=-0.02243909$.

\section{Reduced Coordinates, Condensed Models, and Physical Reconstruction}
\label{app:scalar-applicability-construction}

For the PMMA--aluminum system, response enrichment and fixed-frequency condensation define the three coupling levels. A matrix pencil denotes the expression $\bm K-\lambda\bm M$ associated with $\bm K\bm x=\lambda\bm M\bm x$. The condensed models update the host response at each detuning, and the PMMA--aluminum nominal models freeze it at $\alpha=0$. The fixed channels define the composition weights used to describe family exchange. Related periodic-wave models use parameter-dependent matrix interpolation~\citep{Mencik2026} or Bloch-mode Green's functions for localized defects~\citep{Dorn2026DefectScattering}.

\begin{table}[!b]
  \centering
  \caption{Model hierarchy. The enriched and condensed coupling models apply to the PMMA--aluminum system. Dimensions count reduced coordinates except for FOM. Localization uses reconstructed physical cell mass. All nominal models use consistent Galerkin projection with fixed recovery.}
  \label{tab:model-guide}
  \setlength{\tabcolsep}{3pt}
  \begin{tabular}{L{0.23\linewidth}L{0.20\linewidth}L{0.49\linewidth}}
    \toprule
    Model & Retained space & Construction and role \\
    \midrule
    FOM & Finite-element space & Bloch-reduced eigenproblem for independent checks. \\
    Parameter ROM & 24 & Reference detuning sweeps in the parameter-enriched space. \\
    Frequency-enriched model & 24 & Host-response space for fixed-frequency condensation. \\
    Condensed coupling levels & 6 & Fully coupled, family-separated, and independent-channel models with common recovery $\bm J(\alpha)$. \\
    Off-axis four-coordinate truncation & $B_1$ and $E_2$ at both sites & Threshold prediction and coupling-strength study. \\
    Frozen crossing pair & Two uncoupled crossing modes & Site preference at the nominal crossing and two-state threshold estimates for the diagonal dimers. \\
    Nominal six-coordinate & $B_1+E$ or $B_2+E$ at both sites & Recovery from the frequency-enriched space in the PMMA--aluminum system and the full finite-element complement in the aluminum--tungsten system. \\
    \bottomrule
  \end{tabular}
\end{table}

\subsection{Localized Basis and Interaction Measures}

Let $\bm\Phi$ contain the $m=3$ mass-orthonormal Bloch-reduced single-defect modes of the PMMA--aluminum system. Reflection eigenvectors resolve doublets within their invariant subspaces~\citep{Sakoda1995,Marchal2012}. The PMMA--aluminum local order is $(B_1,E_+,E_-)$, with $E_\pm$ even and odd under $x$-axis reflection, respectively. Translating the modes to both sites gives the $\Gamma$-point basis and operators
\begin{equation}
 \begin{aligned}
  \bm W&=[\bm\Phi_A,\bm\Phi_B],\\
  \bm S&=\bm W^{\mathrm H}\bm M\bm W,\\
  \bm H&=\bm W^{\mathrm H}\bm K\bm W,
 \end{aligned}
 \label{eq:localized-basis}
\end{equation}
where $\bm K,\bm M$ are the Bloch-reduced dimer matrices. Evanescent overlap makes $\bm S$ nonidentity~\citep{Escalante2013}, and its off-diagonal terms can affect mode veering~\citep{JacquesAdhikari2025}. For the positive-definite site blocks $\bm G_X=\bm\Phi_X^{\mathrm H}\bm M\bm\Phi_X$, local mass orthonormalization (whitening) uses
\begin{equation}
 \begin{aligned}
  \bm T_{\mathrm{loc}}&=\operatorname{diag}(\bm G_A^{-1/2},\bm G_B^{-1/2}),\\
  \bm S_\mu&=\bm T_{\mathrm{loc}}^{\mathrm H}\bm S\bm T_{\mathrm{loc}}.
 \end{aligned}
 \label{eq:local-mass-whitening}
\end{equation}

Principal Hermitian inverse roots normalize the two site blocks while preserving their cross-site overlap; $\mu$ denotes this frame.

With $\bm A_{\mathrm{ref}}=\operatorname{diag}(\lambda_1,\ldots,\lambda_m,\lambda_1,\ldots,\lambda_m)$ repeating the single-defect eigenvalues, define
\begin{equation}
 \begin{aligned}
  \bm C_{\mathrm{raw}}&=\bm H-\tfrac12(\bm A_{\mathrm{ref}}^{\mathrm H}\bm S+\bm S\bm A_{\mathrm{ref}}),\\
  \bm C_\mu&=\bm T_{\mathrm{loc}}^{\mathrm H}\bm C_{\mathrm{raw}}\bm T_{\mathrm{loc}}.
 \end{aligned}
 \label{eq:centered-operator}
\end{equation}

Its cross-site block $\bm C_{AB}^\mu$ defines coupling measures and nominal channel directions; the eigenproblem instead uses the condensed stiffness in Eq.~\eqref{eq:fixed-shift-downfolding}. A stiffness-like interaction $c$ (angular frequency squared) is converted to the linearized frequency scale
\begin{equation}
 c^{(f)}=\frac{c}{8\pi^2f_{\mathrm{ref}}},\qquad
 f_{\mathrm{fam}}=(2\pi)^{-1}\sqrt{m^{-1}\sum_{s=1}^m\lambda_s}.
 \label{eq:first-order-equivalent-frequency}
\end{equation}

Here $f_{\mathrm{ref}}>0$ is the single-defect-channel frequency for symmetry-resolved measures and the retained-set RMS frequency $f_{\mathrm{fam}}$ otherwise, providing a first-order frequency scale for comparing coupling strengths.

\subsection{Response Enrichment and Host Condensation}
\label{sec:scalar-applicability}
The constructions combine component-mode reduction and condensation~\citep{CraigBampton1968,KrattigerHussein2014,HagosHan2025} with rational Krylov response enrichment~\citep{Ruhe1984,Bai2002,Benner2015,Lee2024ParametricROM}:
\begin{equation}
 [\bm K_0-\lambda_{\mathrm{enr}}\bm M]\bm X_\ell=\bm B\bm V_{\ell-1},
 \qquad \lambda_{\mathrm{enr}}=(2\pi f_{\mathrm{enr}})^2>0.
 \label{eq:host-response-enrichment}
\end{equation}

Frequency enrichment uses $\bm B=\bm M$ and the translated modal seed $\bm V_0$; PMMA--aluminum parameter enrichment uses $\bm B=\bm K_B$ and nominal coupled Ritz vectors. For each block index $\ell\ge1$, the raw response block $\bm X_\ell$ generated from the previous orthonormalized block $\bm V_{\ell-1}$ is mass-orthogonalized twice against the existing basis, mass-orthonormalized to form $\bm V_\ell$, and used to generate the next block (see Sec.~\ref{app:rom-settings} for seeds, dimensions, and shifts).

The frequency-enriched basis $\bm V_f=[\bm Q_0,\bm Q_h]$, with $\bm Q_0=\bm W\bm S^{-1/2}$ and mass-orthogonal response directions $\bm Q_h$, satisfies $\bm M_f=\bm V_f^{\mathrm H}\bm M\bm V_f=\bm I$. Partitioning $\bm H_f=\bm V_f^{\mathrm H}(\bm K_0+\alpha\bm K_B)\bm V_f$ accordingly, fixed-frequency condensation~\citep{Guyan1965,Friswell1995} gives
\begin{equation}
 \bm H_{\mathrm{eff}}(\alpha;\lambda_{\mathrm{cond}})
 =\bm H_{00}-\bm H_{0h}(\bm H_{hh}-\lambda_{\mathrm{cond}}\bm I)^{-1}\bm H_{h0}.
 \label{eq:fixed-shift-downfolding}
\end{equation}

The shift $\lambda_{\mathrm{cond}}=(2\pi f_{\mathrm{cond}})^2>0$ need not equal $\lambda_{\mathrm{enr}}$. Shifted operators must be nonsingular; condensation additionally requires the response block's smallest-to-largest singular-value ratio to exceed $10^{-14}$ and the solve to converge. The locally whitened pencil uses
\begin{equation}
 \bm K_{\mathrm{eff}}^\mu
 =\bm T_{\mathrm{loc}}^{\mathrm H}(\bm S^{1/2})^{\mathrm H}
 \bm H_{\mathrm{eff}}\bm S^{1/2}\bm T_{\mathrm{loc}},
 \label{eq:metric-downfolded-stiffness}
\end{equation}
with mass $\bm S_\mu$. Physical normalization and defect mass instead use the common recovery in Sec.~\ref{app:physical-reconstruction}.

\subsection{Fixed Channels and Composition Weights}
\label{app:fixed-channel-construction}
In the PMMA--aluminum local order, let $\bm E_{B_1}^{\mathrm{loc}}=\bm e_1$ and $\bm E_E^{\mathrm{loc}}=[\bm e_2,\bm e_3]$, where $\bm e_i\in\mathbb C^3$ are canonical vectors. Their whitened family bases are $\bm Z_{\Gamma,X}=\operatorname{orth}(\bm G_X^{1/2}\bm E_\Gamma^{\mathrm{loc}})$ for $\Gamma\in\{B_1,E\}$ and $X\in\{A,B\}$; $\operatorname{orth}$ returns an orthonormal column basis. The nominal SVD
\begin{equation}
 \bm Z_{E,A}^{\mathrm H}\bm C_{AB}^\mu(0)\bm Z_{E,B}
 =\bm U_E\operatorname{diag}(J_{E,1},J_{E,2})\bm V_E^{\mathrm H},
 \qquad J_{E,1}\geq J_{E,2}\geq0,
 \label{eq:doublet-coupling-svd}
\end{equation}
pairs the directions $\bm Z_{E,A}\bm u_{E,j}$ and $\bm Z_{E,B}\bm v_{E,j}$, where $\bm u_{E,j},\bm v_{E,j}$ are the singular-vector columns. Embedding these sitewise after the $B_1$ pair defines $\bm Z_c^0$ in the channel order of main-text Eq.~\eqref{eq:fixed-channel-coordinates}. This frame is held fixed during detuning; its channels need not be invariant under the full dimer operator. The channel matrices are
\begin{equation}
 \begin{aligned}
  \bm K_c(\alpha)&=(\bm Z_c^0)^{\mathrm H}\bm K_{\mathrm{eff}}^\mu(\alpha)\bm Z_c^0,\\
  \bm S_c&=(\bm Z_c^0)^{\mathrm H}\bm S_\mu\bm Z_c^0.
 \end{aligned}
 \label{eq:fixed-frame-channel-stiffness}
\end{equation}

For a $6\times6$ matrix $\bm X$, let $\mathcal B_{2+4}(\bm X)$ retain its singlet and complete-doublet blocks and $\mathcal B_{2+2+2}(\bm X)$ its three separate two-site channel blocks. The coupling levels in main-text Table~\ref{tab:coupling-levels} are
\begin{equation}
  \begin{aligned}
  k=0:\quad&(\bm{K}^{(0)},\bm{S}^{(0)})=(\bm{K}_c,\bm{S}_c),\\
  k=1:\quad&\bm{K}^{(1)}=\mathcal{B}_{2+4}(\bm{K}_c),\quad
             \bm{S}^{(1)}=\mathcal{B}_{2+4}(\bm{S}_c),\\
  k=2:\quad&\bm{K}^{(2)}=\mathcal{B}_{2+2+2}(\bm{K}_c),\quad
             \bm{S}^{(2)}=\mathcal{B}_{2+2+2}(\bm{S}_c).
  \end{aligned}
  \label{eq:nested-mechanism-models}
\end{equation}

The independent-channel model solves the principal $2\times2$ channel pencils. For the site-pair selector $\bm E_j\in\mathbb C^{6\times2}$, let $\bm x_{j,s}\in\mathbb C^2$, $s=1,2$, denote the eigenvectors of $(\bm E_j^{\mathrm H}\bm K_c\bm E_j,\bm E_j^{\mathrm H}\bm S_c\bm E_j)$.

For $\bm W_\mu=\bm W\bm T_{\mathrm{loc}}$ and $\bm Z_X=[\bm Z_{B_1,X},\bm Z_{E,X}]$, the joint mass projection of a multimode Ritz vector gives
\begin{equation}
 \begin{aligned}
  \begin{bmatrix}\bm{c}_{A,r}\\\bm{c}_{B,r}\end{bmatrix}
    &=\bm{S}_{\mu}^{-1}\bm{W}_{\mu}^{\mathrm{H}}\bm{M}\bm{q}_r,\\
  \bm{a}_r&=\bm{Z}_A^{-1}\bm{c}_{A,r},\\
  \bm{b}_r&=\bm{Z}_B^{-1}\bm{c}_{B,r}.
 \end{aligned}
 \label{eq:metric-symmetry-mode-coordinates}
\end{equation}

The vectors $\bm c_{X,r}$ are locally whitened site coefficients, while $\bm a_r,\bm b_r$ are family coefficients. Writing $\bm a_r=(a_{r,B_1},\bm a_{r,E}^{\mathrm T})^{\mathrm T}$ and similarly for $\bm b_r$, the channel weights are
\begin{equation}
  \begin{aligned}
    \widetilde w_{r,B_1}
    &=|a_{r,B_1}|^2+|b_{r,B_1}|^2,\\
    \widetilde w_{r,E_j}
    &=|\bm{u}_{E,j}^{\mathrm{H}}\bm{a}_{r,E}|^2
      +|\bm{v}_{E,j}^{\mathrm{H}}\bm{b}_{r,E}|^2,
      \qquad j\in\{1,2\}.
  \end{aligned}
  \label{eq:two-level-channel-weights}
\end{equation}

For a nonzero projection, $w_{r,j}=\widetilde w_{r,j}/\sum_k\widetilde w_{r,k}$, where $k\in\{B_1,E_1,E_2\}$. These are projected coefficient weights, not physical modal-mass fractions; their largest value identifies the dominant channel. The site--family weights in main-text Fig.~\ref{fig:offaxis-wrong-defect} sum both doublet components at each site:
\begin{equation}
  \begin{aligned}
    w_{A,B_1}&=|a_{r,B_1}|^2/Z_r,&
    w_{B,B_1}&=|b_{r,B_1}|^2/Z_r,\\
    w_{A,E}&=\lVert\bm{a}_{r,E}\rVert_2^2/Z_r,&
    w_{B,E}&=\lVert\bm{b}_{r,E}\rVert_2^2/Z_r.
  \end{aligned}
  \label{eq:site-family-coefficients}
\end{equation}

Here $\|\cdot\|_2$ is the Euclidean norm, and $Z_r=|a_{r,B_1}|^2+|b_{r,B_1}|^2+\lVert\bm a_{r,E}\rVert_2^2+\lVert\bm b_{r,E}\rVert_2^2>0$. These four weights sum to unity in the projected coefficient space, and summing the two doublet components makes $w_{A,E}$ and $w_{B,E}$ invariant to unitary rotations within the $E$ subspace.

\begin{sloppypar}
Individual $E$ channels are resolved when $g_E>10^{-6}$, where $g_E=|J_{E,1}-J_{E,2}|/\max(J_{E,1},10^{-30})$. All PMMA--aluminum nominal gaps exceed 0.14. Frames constructed at $\alpha=\pm0.01$ and held fixed were compared with the nominal frame at $\alpha=-0.03$, $-0.01$, $0$, $0.01$, $0.03$. Mass-matched independent-channel predictions differ in relative frequency by at most $2.46\times10^{-7}$, with minimum localized-mass MAC 0.999997.
\end{sloppypar}

The relative cross-family interaction is
\begin{equation}
  \rho_{\times}=\frac{1}{\lVert\bm{C}_{AB}^{\mu}\rVert_{\mathrm{F}}}
  \left(
  \lVert\bm{Z}_{B_1,A}^{\mathrm{H}}\bm{C}_{AB}^{\mu}\bm{Z}_{E,B}\rVert_{\mathrm{F}}^2
  +\lVert\bm{Z}_{E,A}^{\mathrm{H}}\bm{C}_{AB}^{\mu}\bm{Z}_{B_1,B}\rVert_{\mathrm{F}}^2
  \right)^{1/2}.
  \label{eq:cross-family-ratio}
\end{equation}

Here $\|\cdot\|_{\mathrm F}$ is the Frobenius norm; set $\rho_\times=0$ for a zero interaction block.

\begin{table}[!b]
  \centering
  \caption{Nominal coupling at $\alpha=0$ for the PMMA--aluminum system. Singular values $\sigma_i^{(f)}$ of $\bm C_{AB}^{\mu}$ use Eq.~\eqref{eq:first-order-equivalent-frequency}; $\rho_\times$ follows Eq.~\eqref{eq:cross-family-ratio}.}
  \label{tab:nominal-coupling}
  \setlength{\tabcolsep}{3pt}
  \begin{tabular}{lccc}
    \toprule
    Metric & Axial $(2,0)$ & Off-axis $(2,1)$ & Diagonal $(2,2)$ \\
    \midrule
    $(\sigma_1^{(f)},\sigma_2^{(f)},\sigma_3^{(f)})$ (Hz) & \begin{tabular}[t]{@{}c@{}}$(8.292,5.178,$\\$2.571)$\end{tabular} & \begin{tabular}[t]{@{}c@{}}$(5.191,3.594,$\\$0.809)$\end{tabular} & \begin{tabular}[t]{@{}c@{}}$(3.436,2.442,$\\$0.372)$\end{tabular} \\
    Cross-family ratio $\rho_{\times}$ & 0.06324 & 0.23450 & 0.39681 \\
    \bottomrule
  \end{tabular}
\end{table}

Table~\ref{tab:nominal-coupling} compares the nominal coupling measures. The axial layout has the largest coupling singular values and the smallest cross-family ratio; these measure interaction strength and family content, respectively.

\subsection{Common Physical Reconstruction}
\label{app:physical-reconstruction}
For a six-coordinate vector $\bm x^{(k)}$ of the three condensed models, including the embedded independent-channel solution $\bm x^{(2)}=\bm E_j\bm x_{j,s}$, response recovery and Bloch reconstruction give
\begin{equation}
 \begin{aligned}
  \bm u^{(k)}&=\bm J(\alpha)\bm x^{(k)},\\
  \bm J(\alpha)&=\bm B_{\mathrm{Bloch}}\bm V_f
 \begin{bmatrix}\bm I\\\bm R(\alpha)\end{bmatrix}
 \bm S^{1/2}\bm T_{\mathrm{loc}}\bm Z_c^0,\\
  \bm R(\alpha)&=-(\bm H_{hh}(\alpha)-\lambda_{\mathrm{cond}}\bm I)^{-1}\bm H_{h0}(\alpha),\\
  \bm Q_X^c(\alpha)&=\bm J(\alpha)^{\mathrm H}\bm M_X^{\mathrm{phy}}\bm J(\alpha).
 \end{aligned}
 \label{eq:app-full-displacement-map}
\end{equation}

Here $\bm B_{\mathrm{Bloch}}$ reconstructs the full physical displacement, $\bm J\in\mathbb C^{N_{\mathrm{phys}}\times6}$, and $N_{\mathrm{phys}}$ is the number of physical degrees of freedom. The matrix $\bm M_X^{\mathrm{phy}}$ contains only elements in defect cell $X\in\{A,B\}$; $\bm Q_X^c\in\mathbb C^{6\times6}$ is its recovered modal-mass form. Both recovery and these forms vary with $\alpha$. At each detuning, the three coupling levels and the target-only control use the same original $\bm J$ and $\bm Q_X^c$; neither is recomputed after deleting coupling blocks. Modes are normalized in the physical mass metric before evaluating $L$.

\section{Branch Continuation, Error Statistics, and Coupling Controls}
\label{app:host-recovery-validation}
The physical recovery in Sec.~\ref{app:physical-reconstruction} permits direct comparison of the localization measures. Independent continuation tests whether reversal persists along a branch, whereas pointwise restoration identifies which omitted interactions dominate the discrepancy at a matched state. The procedures and their supporting checks are specified below.

\subsection{Mode Tracking and Cluster Comparisons}
\label{app:tracking-cluster-metrics}
In the PMMA--aluminum comparisons, independent-channel and reference modes are paired nominally and continued independently in both detuning directions~\citep{KimKim2000,Lu2020,SunKhandelwal2025,Lee2024ParametricROM,Eldred1995}. Hungarian assignment~\citep{Kuhn1955} maximizes the total mass-weighted MAC between successive Bloch-reduced states:
\begin{equation}
  \operatorname{MAC}_M(\bm q_i,\bm q_j)=
  \frac{|\bm q_i^{\mathrm H}\bm M\bm q_j|^2}
  {(\bm q_i^{\mathrm H}\bm M\bm q_i)(\bm q_j^{\mathrm H}\bm M\bm q_j)}.
  \label{eq:physical-mac}
\end{equation}

Physical-space matching uses reconstructed vectors and $\bm M^{\mathrm{phy}}$.

Before assignment, clusters are formed by transitive closure of the adjacent-frequency-gap condition $|f_{i+1}-f_i|/[0.5(f_{i+1}+f_i)]\leq g_{\mathrm{cl}}$.

Independent-channel and reference clusters are detected independently, transferred through the one-to-one mass-correlation assignment, and united. Individual errors are reported only outside this union. The relative gap threshold is $g_{\mathrm{cl}}=10^{-4}$, with sensitivity checks at $5\times10^{-5}$ and $2\times10^{-4}$.

Let $\bm Q_{\mathrm F}$ and $\bm Q_{\mathrm R}$ be mass-orthonormal bases for a pair of compared clusters (independent-channel versus reference, or ROM versus FOM). For equal ranks, the maximum principal angle is $\theta_{\max}=\arccos[\sigma_{\min}(\bm Q_{\mathrm F}^{\mathrm H}\bm M\bm Q_{\mathrm R})]$~\citep{KnyazevArgentati2002}, where $\sigma_{\min}$ denotes the smallest singular value, while rank mismatches are assigned $90^\circ$. Within any cluster, the localization spectrum is defined by
\begin{equation}
  \bm{Q}^{\mathrm{H}}(\bm{M}_A-\bm{M}_B)\bm{Q}\bm{v}_i
  =\ell_i^{\mathrm{cl}}
  \bm{Q}^{\mathrm{H}}(\bm{M}_A+\bm{M}_B)\bm{Q}\bm{v}_i.
  \label{eq:cluster-polarization-spectrum}
\end{equation}

\begin{sloppypar}
Here $\bm Q$ is a Bloch-reduced cluster basis, $\bm v_i$ its state coefficients, and $\bm M_X=\bm B_{\mathrm{Bloch}}^{\mathrm H}\bm M_X^{\mathrm{phy}}\bm B_{\mathrm{Bloch}}$ for $X\in\{A,B\}$. The projected denominator matrix must satisfy $\lambda_{\min}>10^{-12}\lambda_{\max}$, where $\lambda_{\min}$ and $\lambda_{\max}$ are its smallest and largest eigenvalues; otherwise it is rejected as singular. The ordered $\ell_i^{\mathrm{cl}}\in[-1,1]$ are invariant under nonsingular cluster-basis changes. For two equal-rank clusters, the error between their ordered independent-channel and reference spectra is $e_{\ell,\infty}^{\mathrm{cl}}=\max_i\allowbreak|\ell_{i,\mathrm{ind}}^{\mathrm{cl}}-\ell_{i,\mathrm{ref}}^{\mathrm{cl}}|$. At $\alpha=0$, these eigenvectors supply a continuation basis inside each cluster.
\end{sloppypar}

Table~\ref{tab:cluster-threshold-audit} shows the off-axis and diagonal results for the PMMA--aluminum system. The axial branches are resolved at all three tested thresholds. The maximum eligible magnitude errors remain 0.03777 axially and 0.90932 off axis. The diagonal maximum is 0.91498 at the lower threshold and 0.87692 at the central and higher thresholds.

The diagonal PMMA--aluminum pointwise maximum at $\alpha=-0.022$ belongs to an independent-channel-triggered union cluster at all three thresholds, with $\theta_{\max}=5.532^\circ$ and $e_{\ell,\infty}^{\mathrm{cl}}=0.01287$ centrally.

\begin{table}[!t]
  \centering
  \caption{Union-cluster sensitivity for the PMMA--aluminum system. $N_{\mathrm{cl}}$ counts clusters, and $N_{\mathrm{supp}}/N$ is the suppressed fraction of mode--detuning comparisons. Angles and localization-spectrum errors are maxima.}
  \label{tab:cluster-threshold-audit}
  \setlength{\tabcolsep}{1.5pt}
  \begin{tabular}{@{}C{0.15\linewidth}C{0.14\linewidth}C{0.12\linewidth}C{0.16\linewidth}R{0.16\linewidth}R{0.17\linewidth}@{}}
    \toprule
    $\bm{d}$ & $g_{\mathrm{cl}}$ & $N_{\mathrm{cl}}$ & $N_{\mathrm{supp}}/N$ & $\theta_{\max}$ (deg) & $e_{\ell,\infty}^{\mathrm{cl}}$ \\
    \midrule
    $(2,1)$ & $5\times10^{-5}$ & 3 & $6/366$ & 18.37 & 0.129 \\
            & $1\times10^{-4}$ & 4 & $8/366$ & 18.37 & 0.129 \\
            & $2\times10^{-4}$ & 11 & $22/366$ & 19.77 & 0.153 \\
    \addlinespace[2pt]
    $(2,2)$ & $5\times10^{-5}$ & 13 & $26/366$ & 44.14 & 0.701 \\
            & $1\times10^{-4}$ & 29 & $58/366$ & 44.14 & 0.701 \\
            & $2\times10^{-4}$ & 55 & $116/366$ & 40.73 & 0.724 \\
    \bottomrule
  \end{tabular}
\end{table}

\subsection{Six-Coordinate Coupling and Localization Measures at Fixed Mode Shapes}
\label{app:sixdim-controls}
All controls retain the six coordinates of main-text Eq.~\eqref{eq:fixed-channel-coordinates}. The family-separated model removes every $B_1$--$E$ stiffness and mass block; the supplementary target-only control removes only $B_1$--$E_2$ off axis or $B_1$--$E_1$ diagonally, together with Hermitian transposes. Other entries and the original physical forms remain unchanged.

Each model is initialized at $\alpha=0$ by maximum normalized overlap with the lower-frequency target-$E$ mode. Six-mode one-to-one continuation uses $\operatorname{MAC}_c$, defined by Eq.~\eqref{eq:physical-mac} with channel vectors and the original constant metric $\bm S_c$. This metric is unchanged when a control modifies its eigenproblem mass and differs from the reconstructed physical MAC. The fully coupled branches have physical MAC exceeding 0.99967 and 0.99982 with their matched parameter-ROM modes off axis and diagonally. The off-axis fully coupled branch remains resolved. The family-separated model's combined-$E$ block preserves identity at crossings with the separated $B_1$ block. Diagonally, the symmetry-isolated $(B_1,E_1)$ block resolves the fully coupled branch identity at the 13 full-spectrum near-cluster samples.

Table~\ref{tab:sixdim-control-summary} gives endpoints and changes in localization index. In the target-only control, off-axis samples $\alpha=-0.0225,-0.023$ are excluded at $g_{\mathrm{cl}}=10^{-4}$. The remaining 58 nonzero samples have $L\in[-0.749615,-0.018976]$; all 30 diagonal nonzero samples are comparable, with $L\in[-0.820528,-0.046591]$. Thus the resolved segments favor $B$ without reversal. The minimum mass eigenvalues are 0.9577 and 0.9637, respectively, and all sampled generalized eigenvalues are positive.

\begin{table}[!t]
  \centering
  \caption{Branch endpoints at $\alpha=-0.03$ and maximum changes in localization index over 61 off-axis and 31 diagonal samples.}
  \label{tab:sixdim-control-summary}
  \setlength{\tabcolsep}{3pt}
  \begin{tabular}{l*{2}{S[group-digits=none,table-format=+1.6,retain-explicit-plus]}}
    \toprule
    Quantity & {Off-axis} & {Diagonal} \\
    \midrule
    $L^{(0)}(-0.03)$ & +0.732654 & +0.713020 \\
    $L^{(1)}(-0.03)$ & -0.749766 & -0.820528 \\
    Target-only $L(-0.03)$ & -0.749615 & -0.820528 \\
    $L_{\mathrm{diag}}^{(6)}(-0.03)$ & +0.729207 & +0.709632 \\
    $\max|L^{(0)}-L_{\mathrm{diag}}^{(6)}|$ & 0.003519 & 0.003400 \\
    \bottomrule
  \end{tabular}
\end{table}

\subsection{PMMA--Aluminum Branch-Error Statistics}
\label{app:primary-branch-errors}

Along each independently continued branch $b$, we evaluate the signed localization error $e_{L,b}^{\mathrm{sgn}}=|L_b^{\mathrm{ind}}-L_b^{\mathrm{ref}}|$ and the magnitude error $e_{|L|,b}=\bigl||L_b^{\mathrm{ind}}|-|L_b^{\mathrm{ref}}|\bigr|$ at matched detuning values, where $\mathrm{ind}$ and $\mathrm{ref}$ denote the independent-channel and reference models. Eligible comparisons follow the clustering and matching rules of main-text Sec.~\ref{sec:detuning-branch-tracking}, and the dominant channel is identified by maximizing $w_{r,j}$ (Sec.~\ref{app:fixed-channel-construction}). An eligible pair is classified as predicting opposite defects when $L_b^{\mathrm{ind}}L_b^{\mathrm{ref}}<0$ with $\min(|L_b^{\mathrm{ind}}|,|L_b^{\mathrm{ref}}|)\geq0.1$.

\begin{table}[!b]
  \centering
  \caption{Independent-channel--ROM errors on common continued branches over $[-0.03,0.03]$, excluding union clusters at $g_{\mathrm{cl}}=10^{-4}$. Channel weights cover the full sweep; relative frequency errors are normalized by $f^{\mathrm{ROM}}$.}
  \label{tab:two-level-errors}
  \setlength{\tabcolsep}{3pt}
  \begin{tabular}{lrrr}
    \toprule
    Metric & \multicolumn{1}{c}{Axial $(2,0)$} & \multicolumn{1}{c}{Off-axis $(2,1)$} & \multicolumn{1}{c}{Diagonal $(2,2)$} \\
    \midrule
    Minimum dominant-channel weight & 0.961 & 0.474 & 0.503 \\
    Eligible mode--detuning comparisons & 366 & 358 & 308 \\
    Maximum relative frequency error (\%) & 0.0210 & 0.286 & 0.200 \\
    Maximum $e_L^{\mathrm{sgn}}$ & 0.0378 & 1.5701 & 1.5442 \\
    Maximum $e_{|L|}$ & 0.0378 & 0.9093 & 0.8769 \\
    \bottomrule
  \end{tabular}
\end{table}

Table~\ref{tab:two-level-errors} summarizes the eligible comparisons. Cluster sensitivity is documented in Sec.~\ref{app:tracking-cluster-metrics}.

Independent FOM comparisons confirm opposite-defect predictions in the non-axial layouts, with none in the axial case. The complete branch responses are provided in Sec.~\ref{app:full-branch-responses}.

\subsubsection{Full-Branch Responses}
\label{app:full-branch-responses}
\begin{figure}[!t]
  \centering
  \includegraphics[width=0.88\linewidth]{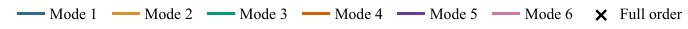}
  \par\smallskip
    \begin{subfigure}[t]{0.44\linewidth}
      \centering
      \includegraphics[width=\linewidth,height=125pt,keepaspectratio]{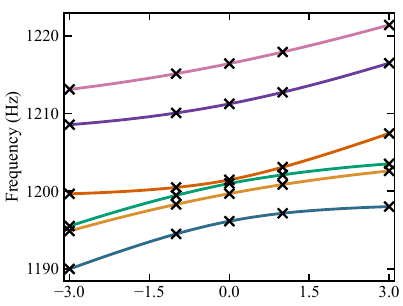}
      \caption{Axial frequency branches.}
    \end{subfigure}
    \hfill
    \begin{subfigure}[t]{0.44\linewidth}
      \centering
      \includegraphics[width=\linewidth,height=125pt,keepaspectratio]{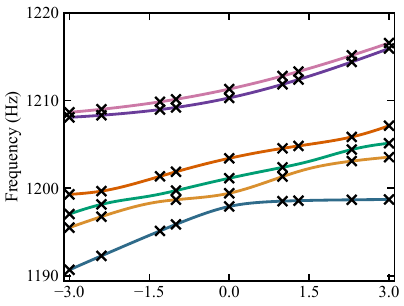}
      \caption{Off-axis frequency branches.}
    \end{subfigure}
    \par\medskip
    \begin{subfigure}[t]{0.44\linewidth}
      \centering
      \includegraphics[width=\linewidth,height=125pt,keepaspectratio]{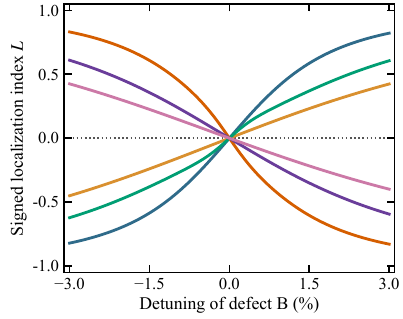}
      \caption{Axial localization.}
    \end{subfigure}
    \hfill
    \begin{subfigure}[t]{0.44\linewidth}
      \centering
      \includegraphics[width=\linewidth,height=125pt,keepaspectratio]{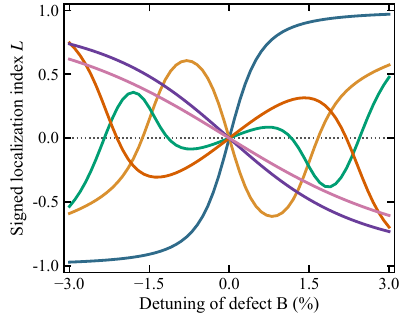}
      \caption{Off-axis localization.}
    \end{subfigure}
  \caption{Parameter-ROM branches under stiffness detuning for the PMMA--aluminum system. In (c,d), positive localization favors $A$ and negative localization favors $B$. Symbols mark independent FOM comparisons.}
  \label{fig:detuning-frequency-polarization}
\end{figure}

Figure~\ref{fig:detuning-frequency-polarization} shows the axial and off-axis frequency and localization branches of the 24-dimensional parameter ROM, with FOM comparisons.

\subsection{Pointwise Restoration and Controlled Error Attribution}
\label{app:pointwise-restoration}
To attribute instantaneous localization errors at a given detuning, each model stage $k$ is matched directly to the parameter ROM at that detuning by maximizing the total physical mass-weighted MAC. For modes paired to the same reference state, the magnitude increments are
\begin{equation}
  \begin{aligned}
  \Delta_{E\mathrm{-mix}}&=|L^{(2)}|-|L^{(1)}|,&
  \Delta_{B_1E}&=|L^{(1)}|-|L^{(0)}|,\\
  \Delta_{\mathrm{base}}&=|L^{(0)}|-|L^{(\mathrm{ROM})}|,&
  \Delta_{\mathrm{ROM}}&=|L^{(\mathrm{ROM})}|-|L^{(\mathrm{FOM})}|.
  \end{aligned}
  \label{eq:mechanism-error-decomposition}
\end{equation}

Here $\Delta_{E\mathrm{-mix}}$ and $\Delta_{B_1E}$ measure the sequential changes upon restoring $E$-channel and $B_1$--$E$ coupling. The increment $\Delta_{\mathrm{base}}$ compares the fully coupled condensed model with the parameter ROM, and $\Delta_{\mathrm{ROM}}$ compares the parameter ROM with the FOM. The first three increments sum to the pointwise independent-channel--ROM magnitude discrepancy. Table~\ref{tab:primary-restoration-summary} summarizes these corrections across all eligible matched states, where the fully coupled model's maximum relative frequency error is $2.10\times10^{-4}$.

Two intermediate models further decompose $\Delta_{\mathrm{base}}$. For $\bm T=[\bm I;\bm R]$ with the recovery of Sec.~\ref{app:physical-reconstruction}, consistent Galerkin projection gives a six-coordinate model with $\bm K_G=\bm T^{\mathrm H}\bm H_f\bm T$ and $\bm M_G=\bm T^{\mathrm H}\bm M_f\bm T$ at $f_{\mathrm{cond}}=\SI{1200}{Hz}$. The uncondensed 24-dimensional frequency-enriched model then solves $\bm H_f\bm v=\lambda\bm M_f\bm v$ directly and is matched to the parameter ROM via $\bm B_f^{\mathrm H}\bm M^{\mathrm{phy}}\bm B_p$, where $\bm B_f$ and $\bm B_p$ are the Bloch-reconstructed physical bases of the frequency- and parameter-enriched spaces.

Out of 1098 modal samples, 1032 remain eligible after cluster exclusion at $g_{\mathrm{cl}}=10^{-4}$, including 195 high-error states with $\bigl||L^{(2)}|-|L^{\mathrm{ROM}}|\bigr|\geq0.1$. Across all 195 high-error pairs, the $B_1$--$E$ restoration exceeds every separately evaluated condensation, restriction, and enrichment correction by at least the minimum margins listed in Table~\ref{tab:primary-restoration-summary}.

\begin{table}[!t]
  \centering
  \caption{Same-state restoration and baseline corrections for the PMMA--aluminum system. The first three rows are maximum localization-magnitude discrepancies or changes over eligible matched states. The last row is the minimum margin by which the $B_1$--$E$ restoration exceeds each separately evaluated baseline correction over the 195 high-error pairs. These margins are evaluated state by state.}
  \label{tab:primary-restoration-summary}
  \setlength{\tabcolsep}{3pt}
  \begin{tabular}{@{}L{0.53\linewidth}*{3}{S[group-digits=none,table-format=1.5]}@{}}
    \toprule
    Quantity & \multicolumn{1}{c}{Axial} & \multicolumn{1}{c}{Off-axis} & \multicolumn{1}{c}{Diagonal} \\
    \midrule
    Fully coupled--parameter ROM maximum error & 0.00730 & 0.02713 & 0.02266 \\
    Maximum change: fully coupled to consistent Galerkin 6D & 0.00033 & 0.01478 & 0.00047 \\
    Maximum change: consistent Galerkin 6D to frequency-enriched 24D & 0.00065 & 0.00040 & 0.00006 \\
    Minimum cross-family-restoration dominance margin & \multicolumn{1}{c}{---} & 0.0939 & 0.1118 \\
    \bottomrule
  \end{tabular}
\end{table}

\FloatBarrier
\end{document}